\documentclass[jcp,twocolumn,amsmath,amssymb]{revtex4-1}

\usepackage[english]{babel}
\usepackage{graphicx,hyperref}
\usepackage{float}
\usepackage{bm}
\usepackage{color}
\usepackage{multirow}

\usepackage{tikz}
\usetikzlibrary{intersections,shapes.arrows}
\usetikzlibrary{shapes.geometric,positioning}

\usepackage{amssymb,amsthm,tikz,fancyhdr,cancel,enumerate,array,booktabs,setspace,pgf,tikz,fancyhdr,graphicx,color}
\usetikzlibrary{patterns,arrows,decorations.pathreplacing}

\definecolor{crimsonglory}{rgb}{0.75, 0.0, 0.2}
\definecolor{lightpastelpurple}{rgb}{0.69, 0.61, 0.85}
\definecolor{mediumpersianblue}{rgb}{0.0, 0.4, 0.65}
\definecolor{ube}{rgb}{0.53, 0.47, 0.76}
\definecolor{seagreen}{rgb}{0.18, 0.55, 0.34}
\definecolor{orange}{rgb}{1, 0.648, 0}
\definecolor{bleuroyal}{rgb}{0.255, 0.41, 0.884}

\begin{document}

\preprint{}

\title{Imaginary Time Correlations for a High-Density two-dimensional Electron Gas}

\author{M. Motta, D.E. Galli}
\affiliation{$\mbox{Dipartimento di Fisica, Universit\`a degli Studi di Milano, via Celoria 16, 20133 Milano, Italy}$}
\author{S. Moroni}
\affiliation{IOM-CNR DEMOCRITOS National Simulation Center and SISSA, via Bonomea 265, 34136 Trieste, Italy}
\author{E. Vitali}
\affiliation{Department of Physics, College of William and Mary, Williamsburg, Virginia 23187-8795, USA}

\date{\today}

\begin{abstract}
We evaluate imaginary time density-density correlation functions for a two-dimensional homogeneous 
electron gas using the phaseless auxiliary field quantum Monte Carlo method.
We show that such methodology, once equipped with suitable numerical stabilization techniques 
necessary to deal with exponentials, products and inversions of large matrices, gives access 
to the calculation of imaginary time correlation functions for medium-sized systems; we present 
simulations of a number up to $42$ correlated fermions in the continuum, using up 
to $300$ plane waves as basis set elements.  
We discuss the numerical stabilization techniques and the computational complexity of the methodology.
We perform the inverse Laplace transform of the obtained density-density correlation functions,
assessing the ability of the phaseless auxiliary field quantum Monte Carlo method to evaluate dynamical 
properties of many-fermion systems.
\end{abstract}
 

\maketitle

\newcommand{\ket}[1]{| #1 \rangle}
\newcommand{\bra}[1]{\langle #1 |}
\newcommand{\braket}[2]{\langle #1 | #2 \rangle}
\newcommand{\tr}[1]{\mbox{tr}\left[ #1 \right]}
\newcommand{\mat}[2]{\mathcal{#1}_{#2}}
\newcommand{\dop}{\hat{\rho}}
\newcommand{\he}{${}^4 \mbox{He}$}
\newcommand{\pvm}[1]{\hat{\Pi}_{#1}}
\newcommand{\hs}{$\mathcal{H}\,$}
\newcommand{\choi}[2]{\ket{#1}\bra{#2}}
\newcommand{\crt}[1]{\hat{a}^\dag_{#1}}
\newcommand{\dst}[1]{\hat{a}_{#1}}
\newcommand{\ham}{\hat{H}}
\newcommand{\evo}[1]{e^{- #1 \ham}}
\newcommand{\gofeta}[1]{\hat{G}(\boldsymbol{\eta}_{#1})}
\newcommand{\gshift}[1]{\hat{G}(\boldsymbol{\eta}_{#1}-\boldsymbol{\xi}_{#1})}
\newcommand{\tns}[3]{\mathcal{#1}^{i_1 \dots i_{#2}}_{\phantom{i_1 \dots i_{#2} \, }j_1 \dots j_{#3}}}
\newcommand{\ster}{\theta,\phi}
\newcommand{\harm}[2]{Y_{#1,#2}(\ster)}
\newcommand{\qzero}{\ket{0}}
\newcommand{\qone}{\ket{1}}
\newcommand{\dt}{\delta\tau}
\newcommand{\psit}{\ket{\Psi_T}}
\newcommand{\eiva}[1]{\epsilon_{#1}}
\newcommand{\eive}[1]{\ket{\Phi_{#1}}}
\newcommand{\evodmc}[1]{e^{- #1 \left(\ham-\eiva{0}\right)}}
\newcommand{\slater}{$\mathfrak{D}(N)$}
\newcommand{\matel}[1]{\langle \hat{#1} \rangle}
\newcommand{\mateldue}[1]{\langle {#1} \rangle}
\newcommand{\vett}[1]{\boldsymbol{#1}}
\newcommand{\rhoq}{\hat{\rho}_{\vett{q}}}
\newcommand{\rhomq}{\hat{\rho}_{-\vett{q}}}

\section{Introduction}

The homogeneous electron gas (HEG) is one of the most widely studied systems in condensed 
matter physics \cite{wigner,bloch,overhauser,ceperley,vignale,shiwei1}.
It represents a model of recognized importance, which offers the opportunity to explore the quantum 
behavior of many-body systems on a fundamental basis and provides a ground test for several quantum 
chemistry \cite{Szabo1996}, many-body \cite{fetter} and quantum Monte Carlo (QMC) \cite{ceperley,
tanatar,kwon,saverio1} methodologies. Furthermore, recent years have witnessed the realization of 
increasingly high-quality two-dimensional (2D) HEGs in devices of considerable experimental interest 
such as quantum-well structures \cite{sperimentali1,sperimentali2} and field-effect transistors
\cite{sperimentali3}. 

The accuracy of QMC calculations for the HEG is unavoidably limited by the well-known sign problem 
\cite{feynman,loh}, arising from the antisymmetry of many-fermion wavefunctions.
The vast majority of QMC simulations of many-fermion systems circumvent the sign problem relying on
the Fixed-Node (FN) approximation \cite{fn,fn2}.
Methodologies based on the FN approximation provide very accurate estimations of ground state properties
such as the kinetic and potential energy, and the static structure factor.
On the other hand, as the extension of the FN approximation to the manifold of excited states is 
less understood and established \cite{noi,Ceperley1991}, the study of dynamical properties of 
many-fermion systems is a very active and challenging research field \cite{fci,assaad,jcp_vari,fc,jcp_dyn,noi,chan1}.

In a recent work \cite{noi}, performing an extensive study of exactly solvable few-fermion Hamiltonians, 
we have shown that the phaseless Auxiliary Fields Quantum Monte Carlo (AFQMC) \cite{scalapino,af1,af3/2,
af2,assaad,senechal,shiwei_bp,shiwei_bp2,jcpshiwei1,jcpshiwei2,jcpshiwei3} method can become an important tool for 
the calculation of imaginary time correlation functions.

Motivated by this result, in the present work we apply the phaseless AFQMC method to the 2D HEG. 
In particular, we focus on the high-density regime ($r_s \leq 2$), since our previous study has revealed 
that the computational cost of the algorithm increases severely with $r_s$, making a study at high $r_s$ 
hardly practicable. This is due to the fact that, increasing $r_s$, the number of plane
waves required to reach convergence in the basis set size becomes larger, making
cumbersome to perform the linear algebra operations required by the methodology. From a more physical
point of view, the stronger correlations give rise to a more pronounced curvature in the wave function,
which is accurately reproduced by a large number of Fourier coefficients. 
Nevertheless, the high-density regime is extremely interesting as the presence of the interaction 
leads to the emergence of important correlation effects, enhanced by the low dimensionality. 
We evaluate density-density correlation functions in imaginary time $F({\vett{q}},\tau)$ and we 
perform their inverse Laplace transform to extract information about the excitations of the system.
We introduce and describe a method for stabilizing the calculation of imaginary time correlation
functions in AFQMC, and present numerical tests demonstrating its accuracy.
We finally assess the accuracy of the calculations comparing AFQMC results with predictions within the 
random phase approximation (RPA) for finite systems \cite{fetter,Sawada1957}. We also compare AFQMC 
estimates with the results of Fixed-Node calculations 
\cite{saverio1,rept}, performed with a nodal structure encompassing optimized rational backflow correlations 
\cite{kwon,Umrigar2008,Motta2015}.

The paper is organized as follows: the phaseless AFQMC method is briefly reviewed in Section 
\ref{method}, the results of the study are discussed in Section \ref{results}, and conclusions are 
drawn in the last Section \ref{conclusions}.

\section{Methodology}
\label{method}

\subsection{The Model}
\label{model}
The 2D HEG is a system of charged spin-$\frac{1}{2}$ fermions interacting with the Coulomb potential
and immersed in a uniform positively-charged background.
For the purpose of studying the 2D HEG we simulate a system of $N$ particles moving inside a square 
region $\mathcal{R}$ of surface $\Omega = L^2$, employing periodic boundary conditions (PBC) at the boundaries of 
the simulation domain, in conjunction with an Ewald summation procedure \cite{ewald}. In the present 
work, energies are measured in Hartree units $E_{Ha}$, and lengths in Bohr radii $a_B$. 
The Hamiltonian of the system reads, in such units:
\begin{equation}
\label{heg_ham}
\ham=\sum_{\vett{k}\sigma} \frac{|\vett{k}|^2}{2} \,\crt{\vett{k}\sigma}\dst{\vett{k}\sigma} +
\frac{1}{2\Omega} \,\sum_{\vett{q}\neq 0} \frac{2\pi}{|\vett{q}|}
\sum_{\substack{\vett{k}\sigma\\\vett{p}\varsigma}}
\crt{\vett{k}+\vett{q}\sigma} \crt{\vett{p}-\vett{q}\varsigma}
\dst{\vett{p}\varsigma} \dst{\vett{k}\sigma}
\end{equation}
where spin-definite plane waves:
\begin{equation}
\label{p_waves}
\varphi_{\vett{k}\sigma}(\vett{r},\omega)
= \frac{e^{i \vett{k}\cdot\vett{r}}}{\sqrt{\Omega}}
\delta_{\omega,\frac{1}{2}-\sigma}
\quad 
\frac{L}{2\pi} \vett{k} \in \mathbb{Z}^2, \,\,
\sigma=\pm \frac{1}{2}
\end{equation}
with $\vett{r} \in \mathcal{R}$, $\omega = 0,1$
are used as a basis for the single-particle Hilbert space. The ground-state 
energy per particle of the system is obtained adding, to the mean value of \eqref{heg_ham}, 
the corrective constant term:
\begin{equation}
\xi = \frac{1}{2L} \left[ 2 
\sum_
{\substack{\vett{n} \in \mathbb{Z}^2 \\ \vett{n} \neq \vett{0}\,\, }}
\frac{ \textrm{erfc}\left( \sqrt{\pi} |\vett{n}| \right) }
     {                                |\vett{n}|         } - 4 \right]
= - 3.900265 \, \frac{1}{2L}
\end{equation}
arising from the Ewald summation procedure employed \cite{ewald}. The Hamiltonian \eqref{heg_ham}
can be parametrized in terms of the dimensionless Seitz radius $r_s$ defined by:
\begin{equation}
\frac{\Omega}{N} = \frac{1}{n} = \pi r_s^2 a_B^2
\end{equation}
where $n$ is the density of the system and $a_B$ the Bohr radius. This parametrization shows that 
the matrix elements of the kinetic energy roughly scale as $|\vett{k}|^2 \simeq r_s^{-2}$, 
and those of the potential energy as $1 / \Omega|\vett{q}| \simeq r_s^{-1}$. 
Thus, for increasing Seitz radius, the interaction part of $\ham$ plays a more and more
relevant role.

%We remind that, unlike some particular models like the negative-$U$ Hubbard model and
%the positive-$U$ Hubbard model in a bipartite lattice \cite{Hirsch1985}, the HEG model is severely
%affected by the fermion sign problem for all values of $r_s$.
%Its simulation by means of configurational or determinantal projective QMC methods is therefore a 
%very delicate and challenging issue \cite{Foulkes2001,fn,fn2,fci}.}

\subsection{The Phaseless AFQMC}
\label{phaseless_afqmc}

To address the calculation of static and dynamical properties of the 2DHEG, we rely on the
phaseless AFQMC method \cite{scalapino,af1,af3/2,af2,assaad,senechal,shiwei_bp,shiwei_bp2,jcpshiwei1,
jcpshiwei2,jcpshiwei3},     that moves from observation that the imaginary time propagator 
$\evo{\tau}$ acts as a projector onto the ground state $\ket{\Phi_0}$ of the system in the 
limit of large imaginary time. Therefore, as long as a trial state $\ket{\Psi_T}$ has non-zero 
overlap with $\ket{\Phi_0}$ the relation:
\begin{equation}
\label{it_prop}
\ket{\Phi_0}
\propto
\lim_{\tau\to\infty} \evodmc{\tau} \psit
\end{equation}
holds, $\epsilon_0$ being the ground state energy, that can be estimated adaptively following a 
common procedure in Diffusion Monte Carlo (DMC) calculations
\cite{Foulkes2001}. 
%\cite{boh}. 
QMC methods rely on the observation that the deterministic evolution \eqref{it_prop} can be 
mapped onto suitable stochastic processes and solved by randomly sampling appropriate probability 
distributions.
Determinantal QMC methods, such as the phaseless AFQMC, use a Slater determinant as trial state 
$\psit$, typically the Hartree-Fock state, and map \eqref{it_prop} onto a stochastic process in 
the abstract manifold \slater $\,$ of $N$-particle Slater determinants. This association is 
accomplished by a discretization of the imaginary time propagator $\evodmc{\tau}$:
\begin{equation}
\label{it_prop_2}
\evodmc{\tau} \psit = \left( \evodmc{\dt} \right)^n \psit
\quad\quad n \in \mathbb{N}, \dt = \frac{\tau}{n}
\end{equation}
and by a combined use of the Trotter-Suzuki decomposition \cite{trotter,suzuki}, of the 
Hubbard-Stratonovich transformation \cite{hubbard,stratonovich,senechal} and of an importance sampling 
technique \cite{senechal,noi} on the propagator $\evodmc{\dt}$. The result is:
\begin{equation}
\label{randomwalk}
\nonumber
\evodmc{\dt} \psit \simeq \int dg(\boldsymbol{\eta}) \, 
\mathfrak{W}\left[ \boldsymbol{\eta}, \boldsymbol{\xi} \right] \, 
\frac{\gshift{} \psit}{\braket{\Psi_T}{\gshift{}|\Psi_T}}\\
\end{equation}
where $dg(\boldsymbol{\eta})$ is a multidimensional standard normal probability distribution,
$\hat{G}(\boldsymbol{\eta})$ is a product of exponentials of one-body operators and:
\begin{equation}
\label{weight}
\mathfrak{W}\left[ \boldsymbol{\eta}, \boldsymbol{\xi} \right] =
e^{-\frac{\boldsymbol{\xi} \cdot \boldsymbol{\xi}}{2} - \boldsymbol{\eta} \cdot \boldsymbol{\xi}} \langle \Psi_T | \hat{G}(\boldsymbol{\eta}-\boldsymbol{\xi}) | \Psi_T \rangle
\end{equation}
is a weight function depending on a complex-valued parameter $\boldsymbol{\xi}$, which is
chosen to minimize fluctuations in $\mathfrak{W}\left[ \boldsymbol{\eta}, \boldsymbol{\xi} \right]$ to first order in $\dt$.
Equation \eqref{randomwalk} illustrates the mechanism responsible for the appearence of
the sign problem in the framework of AFQMC:
\emph{when the overlap between $\hat{G}(\boldsymbol{\eta}-\boldsymbol{\xi}) | \Psi_T \rangle$
and the trial state vanishes} massive fluctuations occur in \eqref{randomwalk}.
In the method conceived by S. Zhang, the {\em{exact}} complex-valued weight function 
appearing in \eqref{weight} is replaced \cite{af2,noi} by the approximate form:
\begin{equation}
\label{rle}
\mathfrak{W}\left[ \boldsymbol{\eta}, \boldsymbol{\xi} \right] \simeq 
e^{-\dt \left(\epsilon_{loc}\left(\hat{G}(\boldsymbol{\eta}-\boldsymbol{\xi})|\Psi_T \rangle\right)-\epsilon_0\right)}
\times
\max(0,\cos(\Delta\theta))
\end{equation}
where $\epsilon_{loc}(\Psi)=\mbox{Re}\left[\frac{\braket{\Psi_T}{\ham|\Psi}}{\braket{\Psi_T}{\Psi}}\right]$ is the \emph{local energy} functional, and:
\begin{equation}
\label{phase_app}
\Delta \theta = \mbox{Im} \left[ \log\left[ \frac{\braket{\Psi_T}{\hat{G}(\boldsymbol{\eta}-\boldsymbol{\xi}) | \Psi \rangle}}{\braket{\Psi_T}{\Psi}} \right] \right]
\end{equation}
The first factor corresponds to the {\em{real local energy approximation}}, which turns 
\eqref{weight} into a {\em{real quantity}}, avoiding phase problems rising from complex 
weights; the real local energy approximation is implemented neglecting some fluctuations 
in the auxiliary fields \cite{noi}.
The second factor, together with the introduction of the shift parameters, has been argued
in \cite{af2,senechal} to keep the overlap between the determinants involved in the random
walk and the trial determinant far from zero.
In fact, the angle $\Delta \theta$ corresponds to the flip in the phase of a determinant
during a step of the random walk: the term $\max(0,\cos(\Delta\theta))$
is meant to suppress determinants whose phase undergoes an abrupt change, under the assumption
\cite{af2,senechal} that such behaviour indicates the vanishing of the overlap with the
trial state.

\subsection{Hubbard-Stratonovich Transformation in the plane-wave Basis Set}
\label{hs_for_pw_basis}
In general, the structure of $\hat{G}(\vett{\eta})$ is specified through a procedure that might result
lengthy and computationally expensive \cite{senechal,noi}. When spin-definite plane waves are used as a 
basis for the single-particle Hilbert space, a remarkable simplification derived in 
details in Appendix \eqref{appA} occurs in its calculation and leads to the following result:
\begin{equation}
\label{hs-transform2}
\hat{G}(\vett{\eta}) = e^{-\frac{\delta\tau}{2} \ham_0}
e^{-i\sqrt{\delta\tau} 
\sum_{\vett{q}\neq0} \eta_{1\vett{q}} \hat{A}_1(\vett{q})+\eta_{2\vett{q}} \hat{A}_2(\vett{q})}
e^{-\frac{\delta\tau}{2} \ham_0}
\end{equation}
with:
\begin{equation}
\label{heg_ham3}
\begin{split}
\ham_0 &= \sum_{\vett{k}\sigma} 
\left( \frac{|\vett{k}|^2}{2} - 
       \frac{1}{2\Omega} \sum_{\vett{p}\neq\vett{k}} \frac{2 \pi}{|\vett{p}-\vett{k}|}
\right)
                         \,\crt{\vett{k}\sigma}\dst{\vett{k}\sigma} = \\
&= \sum_{\vett{k}\sigma} \left( \mathcal{H}_0 \right)_{\vett{k}} \,\crt{\vett{k}\sigma}\dst{\vett{k}\sigma}
\end{split}
\end{equation}
and, denoting $\rhoq$ the Fourier component of the local density:
\begin{equation}
\label{heg_ham4}
\hat{A}_1(\vett{q}) = \sqrt{\frac{2 \pi}{\Omega |\vett{q}|}} \, \frac{\rhoq+\rhomq}{2}
\quad
\hat{A}_2(\vett{q}) = \sqrt{\frac{2 \pi}{\Omega |\vett{q}|}} \, \frac{i\rhoq-i\rhomq}{2}
\end{equation}
The operators \eqref{heg_ham4} will be henceforth written as:
\begin{equation}
\hat{A}_s(\vett{q}) = \sum_{\vett{k}\vett{p}\sigma} \left( \mathcal{A}_s(\vett{q}) 
\right)_{\vett{k}\vett{p}} \, \crt{\vett{k}\sigma}\dst{\vett{p}\sigma}
\end{equation}
with:
\begin{equation}
\begin{split}
\left( \mathcal{A}_1(\vett{q}) \right)_{\vett{k}\vett{p}} = 
\sqrt{\frac{2 \pi}{\Omega |\vett{q}|}} \, \frac{ \delta_{\vett{k},\vett{p}+\vett{q}} + 
                                                 \delta_{\vett{k},\vett{p}-\vett{q}} }{2} \\
\left( \mathcal{A}_2(\vett{q}) \right)_{\vett{k}\vett{p}} = 
\sqrt{\frac{2 \pi}{\Omega |\vett{q}|}} \, \frac{i \delta_{\vett{k},\vett{p}+\vett{q}} - i
                                                 \delta_{\vett{k},\vett{p}-\vett{q}} }{2} \\
\end{split}
\end{equation}
We remind that formulae \eqref{hs-transform2},\eqref{heg_ham3} and \eqref{heg_ham4} result 
from an exact calculation, immediately generalizable to all radial two-body interaction potentials,
and to all  spatial dimensionalities.

\subsection{Numeric Implementation}
\label{phaseless_afqmc_num} 
\begin{figure}
\includegraphics[width=0.4\textwidth]{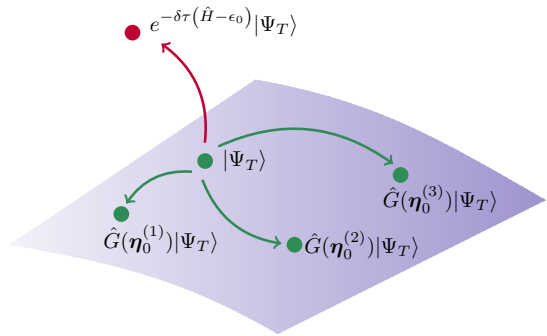}
\caption{
(color online) pictorial representation of the random walk in the manifold of 
$N$-particle Slater determinants $\mathfrak{D}(N)$ (lavender surface). 
The figure points out that the imaginary time propagator $e^{-\delta\tau \left(
\hat{H}-\epsilon_0\right)}$ drives a Slater determinant $\ket{\Psi_T}$ away 
from $\mathfrak{D}(N)$, while the one-body propagators $\hat{G}({\mathbf{\eta}})$
preserve $\mathfrak{D}(N)$. 
This permits to retrieve the analytically intractable state $e^{-\delta\tau 
\left(\hat{H}-\epsilon_0\right)} \ket{\Psi_T}$ as a stochastic linear combination 
of Slater determinants $\hat{G}({\mathbf{\eta}}^{(w)}_0)\ket{\Psi_T}$
according to \eqref{estimate}.}
\label{fig:surface}
\end{figure}

The observations outlined above give rise to a \emph{polynomially complex} algorithm
for numerically sampling \eqref{it_prop}, a pictorial representation of which is given in Fig.~\ref{fig:surface}.
Several Slater determinants $\{ \ket{\Psi^{(w)}_0} \}_{w=1}^{N_w}$, called \emph{walkers}, are initialized to the 
Hartree-Fock ground state, a filled Fermi sphere in the case of translationally invariant systems such as the 2D
HEG, and given initial weights $\{\mathfrak{W}^{(w)}_{0} \}_{w=1}^{N_w} $ equal to $1$.
%\begin{center}
%\end{center}

Subsequently, each walker is let evolve under the action of the operators 
\eqref{hs-transform2} and its weight is updated through multiplication by \eqref{rle}.
An estimate for the ground state of the system is provided by the following stochastic linear
combination of Slater determinants:
\begin{equation}
\label{estimate}
\ket{\Phi_0} \simeq
\frac{ 1 }{ \sum_{w=1}^{N_w} \mathfrak{W}^{(w)}_{n} }
\sum_{w=1}^{N_w} \mathfrak{W}^{(w)}_{n} \frac{\ket{\Psi^{(w)}_{n}}}{\braket{\Psi_T}{\Psi^{(w)}_{n}}}
\end{equation}
Since numeric calculations can be carried out on finitely-generated Hilbert spaces only,
the numeric implementation of the phaseless AFQMC algorithm requires the single-particle Hilbert space 
basis \eqref{p_waves} of the system to be truncated, i.e. only the $M$ lowest-energy plane-waves to
be retained.

\subsection{Dynamical Correlation Functions}
\label{afqmc_dyn_corr}
 
The formalism outlined in \ref{phaseless_afqmc}, \ref{phaseless_afqmc_num} enables the calculation of
ground state properties, and also of the \emph{imaginary time correlation function} (ITCF):
\begin{equation}
\label{ITCF}
\begin{split}
F_{\hat{A},\hat{B}}(\tau) &= 
\braket{\Phi_0}{ \hat{A} \evodmc{\tau} \hat{B} | \Phi_0} \\
\end{split}
\end{equation}
of two single-particle many-body operators $\hat{A},\hat{B}$. \eqref{ITCF} is a purely
mathematical function, related to the \emph{dynamical} or \emph{energy-resolved structure 
factor}:
\begin{equation}
\label{structure}
S_{\hat{A},\hat{B}}(\omega) = \int_{\mathbb{R}} dt \, \frac{e^{i\omega t}}{2 \pi} \, \braket{\Phi_0}{ \hat{A}(t)\hat{B} | \Phi_0}
\end{equation}
of $\hat{A},\hat{B}$, a quantity appearing in linear response theory and providing
precious information on the time-dependent response of the system to external fields.
Dynamical structure factors and ITCFs are related to each other, as revealed by 
their Lehmann representation, by a Laplace transform \cite{fetter}.
% \cite{boh}.
%Such analytic continuation has to be performed numerically, starting from an average of 
%QMC calculations of $F_{\hat{A},\hat{B}}(\tau)$, each affected by statistical noise, in 
%correspondence with a finite number of imaginary time values.
%In the present work, this issue is faced relying on the recently established PIFT
%methodology\cite{fc}.

Within the AFQMC formalism, the issue of computing ITCFs is complicated by the circumstance that the 
one-body operators $\hat{A},\hat{B}$ do not map Slater determinants onto Slater determinants, but on 
rather complicated states. 
Nevertheless, making use of the canonical anticommutation relations between fermionic creation and 
destruction operators it is possible to show \cite{noi} that:
\begin{equation}
\label{swap_op}
\evodmc{\dt} \hat{B} = \int dg(\boldsymbol{\eta}) \, 
\hat{B}(\boldsymbol{\eta}) \hat{G}(\boldsymbol{\eta})
\end{equation}
where $\hat{B}(\boldsymbol{\eta})$ is a suitable one-particle operator. In the case of the 2D HEG, 
it reads:
\begin{equation}
\begin{split}
\hat{B}(\boldsymbol{\eta}) &= \sum_{\vett{k}\vett{p}\sigma} 
\left( \mathcal{B}(\boldsymbol{\eta}) \right)_{\vett{k}\vett{p}} \, 
\crt{\vett{k}\sigma}\dst{\vett{p}\sigma}
\end{split}
\end{equation}
where:
\begin{equation}
\mathcal{B}(\boldsymbol{\eta}) = \mathcal{D}(\boldsymbol{\eta}) \mathcal{B} \mathcal{D}(\boldsymbol{\eta})^{-1}
\end{equation}
is defined through:
\begin{equation}
\begin{split}
\left( \mathcal{D}(\boldsymbol{\eta}) \right)_{\vett{k}\vett{p}} = 
e^{-\frac{\delta\tau}{2} \left( \mathcal{H}_0 \right)_{\vett{k}} } 
\left( e^{- i \sqrt{\dt} \sum_{\vett{q}s} \eta_{\vett{q}s} 
\mathcal{A}_s(\vett{q}) } \right)_{\vett{k}\vett{p}}
e^{-\frac{\delta\tau}{2} \left( \mathcal{H}_0 \right)_{\vett{p}} }
\end{split}
\end{equation}
By application of \eqref{swap_op} and of the backpropagation technique  \cite{shiwei_bp,shiwei_bp2,noi}, it is 
possible to express the ITCF $F_{\hat{A},\hat{B}}(\tau)$ as mean value of a random variable over the 
random path followed by the walkers in the manifold of Slater determinants.

\begin{figure*}
\includegraphics[width=0.65\textwidth]{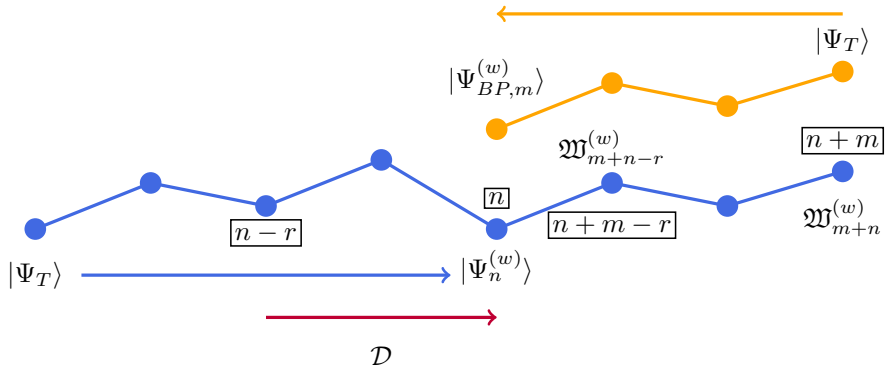}
\caption{(color online) pictorial representation of the phaseless AFQMC estimator for $F_{\hat{A},\hat{B}}(\tau)$, equation 
\eqref{the_boss}; $F_{\hat{A},\hat{B}}(\tau)$ is computed at $\tau = r \, \delta\tau$ with $r=2$, with $n=5$ propagation steps 
and $m=3$ backpropagation steps. The matrix $\mathcal{D}$ appearing in \eqref{dmatrix} is computed between the time steps 
$n-r$ and $n$ (at which $\ket{\Psi^{(w)}_{n}}$ is computed); the determinant $\ket{\Psi^{(w)}_{BP,m}}$ is computed between the 
time steps $n$ and $n+m$, and the weights $\mathfrak{W}^{(w)}_{m+n-r}$ and $\mathfrak{W}^{(w)}_{m+n}$ are computed at the time 
steps $m+n-r$ and $m+n$.} \label{fig:snake}
\end{figure*}

Further details of this calculation procedure are reported in \cite{noi}. For the purpose of the present
work, it is sufficient to recall that the phaseless AFQMC estimator of $F_{\hat{A},\hat{B}}(\tau)$ reads:
\begin{equation}
\label{the_boss}
\begin{split}
F_{\hat{A},\hat{B}}(r \delta\tau) \simeq & \frac{1}{\sum_{w=1}^{N_w} \mathfrak{W}^{(w)}_{m+n-r}} \\
&\sum_{w=1}^{N_w} \sum_{ijkl} \, \mathcal{B}_{kl} \, \mathfrak{W}^{(w)}_{m+n} \, \frac{\langle \Psi^{(w)}_{BP,m}|\hat{A} \hat{a}^\dag_i \hat{a}_j|\Psi^{(w)}_{n} \rangle}{\langle \Psi^{(w)}_{BP,m}|\Psi^{(w)}_n \rangle} \\
&\mathcal{D}(\boldsymbol{\eta}^{(w)}_{n-1} - \boldsymbol{\xi}^{(w)}_{n-1}, \dots , \boldsymbol{\eta}^{(w)}_{n-r} - \boldsymbol{\xi}^{(w)}_{n-r} )_{ik} \\
&\mathcal{D}^{-1}( \boldsymbol{\eta}^{(w)}_{n-1} - \boldsymbol{\xi}^{(w)}_{n-1}, \dots , \boldsymbol{\eta}^{(w)}_{n-r} - \boldsymbol{\xi}^{(w)}_{n-r} )_{lj} \\
\end{split}
\end{equation}
where:
\begin{equation}
\begin{split}
| \Psi^{(w)}_{BP,m} \rangle &= \hat{G}^\dag(\boldsymbol{\eta}_n-\boldsymbol{\xi}_n) \dots \hat{G}^\dag(\boldsymbol{\eta}_{n+m-1}-\boldsymbol{\xi}_{n+m-1}) | \Psi_T \rangle \\
\end{split}
\end{equation}
and:
\begin{equation}
\label{dmatrix}
\begin{split}
  &\mathcal{D}(\boldsymbol{\eta}^{(w)}_{n-1} - \boldsymbol{\xi}^{(w)}_{n-1}, \dots , \boldsymbol{\eta}^{(w)}_{n-r} - \boldsymbol{\xi}^{(w)}_{n-r} ) = \\
= &\mathcal{D}(\boldsymbol{\eta}^{(w)}_{n-1} - \boldsymbol{\xi}^{(w)}_{n-1}) \dots \mathcal{D}( \boldsymbol{\eta}^{(w)}_{n-r} - \boldsymbol{\xi}^{(w)}_{n-r} )
\end{split}
\end{equation}
The estimator \eqref{the_boss} is essentially a weighted average of suitably-constructed matrix elements; each walker 
$w$ constructs the matrix element and the weights $\mathfrak{W}^{(w)}_{m+n-r}$, $ \mathfrak{W}^{(w)}_{m+n}$ involved 
in the weighted average \eqref{the_boss} from two Slater determinants $\ket{\Psi^{(w)}_{n}}$, $\ket{\Psi^{(w)}_{BP,m}}$ 
and two matrices $\mathcal{D}(\boldsymbol{\eta}^{(w)}_{n-1}-\boldsymbol{\xi}^{(w)}_{n-1},\dots,\boldsymbol{\eta}^{(w)}_{n-r}
-\boldsymbol{\xi}^{(w)}_{n-r} )$
These objects are functions of the auxiliary fields configurations $\boldsymbol{\eta}^{(w)}$ defining the random path 
followed by the walker in the manifold of Slater determinants, and their calculation is pictorially illustrated in 
Fig.~\ref{fig:snake}.

In the present work, we consider the imaginary time density-density correlation function:
\begin{equation}
\label{fqt}
F({\vett{q}},\tau) = \frac{\braket{\Phi_0}{ \rhomq \evodmc{\tau} \rhoq | \Phi_0}}{N}
\end{equation}
which is the Laplace transform of the dynamical structure factor $S(\vett{q},\omega)$. This quantity is 
notoriously related to the differential cross section of electromagnetic radiation scattering, and provides 
essential information for the quantitative description of
excitations of the HEG, collective charge density flutuations, i.e. plasmons,
end electron-hole excitations \cite{fetter,vignale}. 

\subsection{Numeric Stabilization}

In a previous work \cite{noi} we have pointed out that, due to the presence of such matrix elements, the
estimator \eqref{the_boss} is negatively-conditioned by a form of numeric instability.
The aim of the present section is to elucidate the origin of such phenomenon and to propose a method for 
stabilizing the calculation of ITCFs in AFQMC.
The AFQMC estimator \eqref{the_boss} of the ITCF $F_{\hat{A},\hat{B}}(\tau)$ involves a weighted average, 
over the random paths followed by the $N_w$ walkers employed in the simulation, of a quantity in which 
the matrix elements of $\mathcal{D}(\boldsymbol{\eta}^{(w)}_{n-1} - \boldsymbol{\xi}^{(w)}_{n-1}, \dots , 
\boldsymbol{\eta}^{(w)}_{n-r} - \boldsymbol{\xi}^{(w)}_{n-r} )$ and of its inverse appear. In the reminder
of the present section, these matrices will be referred to as $\mathcal{D}$ and $\mathcal{D}^{-1}$ for brevity.
The matrices $\mathcal{D}$ and $\mathcal{D}^{-1}$ need to be computed numerically, respectively as 
product of $r$ matrices and inverse of $\mathcal{D}$. It is well-known that the numerical computation of 
$\mathcal{D}$ and $\mathcal{D}^{-1}$ introduces rounding-off errors \cite{Turing1948}, which accumulate 
as $r$ increases with detrimental impact on the results of the computation \cite{loh}.

\begin{figure}
\hspace*{-3cm}\includegraphics[width=0.85\textwidth]{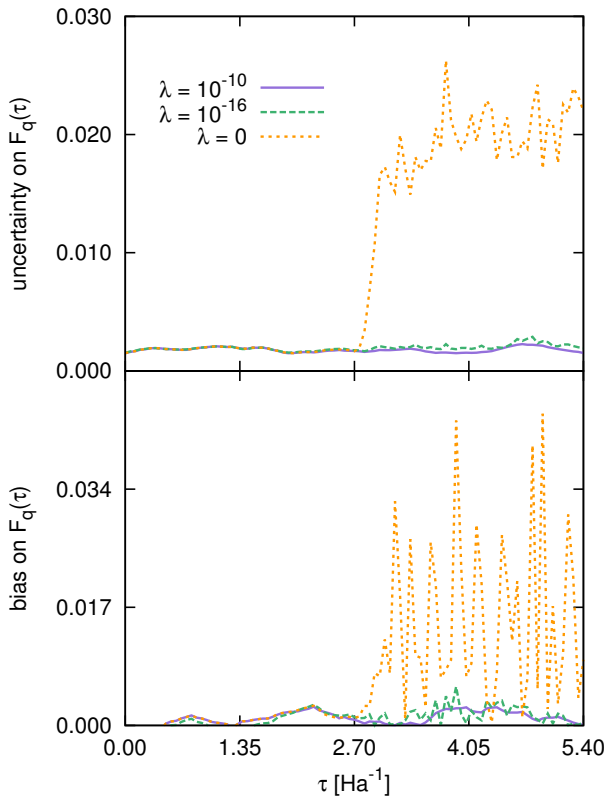}
\caption{(color online) Effect of the Tykhonoff regularization \eqref{eq:tyko} on an ITCFs 
relative to a system of $N=2$ electrons with $M=21$ basis set elements. 
Upper panel: statistical uncertainty affecting the AFQMC estimate of $F(q,\tau)$ with $\lambda = 10^{-10}$ 
(lavender solid lines), $\lambda=10^{-16}$ (green dashed lines) and $\lambda = 0$ (orange    
dotted lines).
Lower panel: bias affecting the AFQMC estimate of $F(q,\tau)$.
%Upper panel: comparison 
%between the exact $F(q,\tau)$ (black line) and the AFQMC estimates with $\lambda = 10^{-10}$ 
%(lavender solid symbols), $\lambda=10^{-16}$ (green dashed symbols) and $\lambda = 0$ (orange 
%dotted symbols).
%Central panel: error bar affecting the AFQMC estimate of $F(q,\tau)$. 
%Lower panel: bias affecting the AFQMC estimate of $F(q,\tau)$.
} \label{fig:tyko}
\end{figure}

Rounding-off errors are particularly severe when the $\infty$-norm condition number:
\begin{equation}
\kappa(\mathcal{D}) = \| \mathcal{D} \|_\infty \| \mathcal{D}^{-1} \|_\infty
\end{equation}
of the matrix $\mathcal{D}$, in which $\| A \|_\infty = \max_{ij} |A_{ij}|$ denotes the
$\infty$-norm on the space of $M \times M$ complex-valued matrices, is large. For the systems under
study, we observe a condition number roughly increasing as $\kappa(\mathcal{D}) \simeq C_1^r$ for some 
constant $C_1$. The rapid increase of $\kappa(\mathcal{D})$ indicates that the numeric matrix inversion
$\mathcal{I}\left( \mathcal{D} \right)$ used to estimate $\mathcal{D}^{-1}$ might be ill-conditioned, 
an intuition that can be confirmed by studying the {\emph figure of merit}:
\begin{equation}
\label{fom}
\| E \|_\infty = \| \mathbb{I} - \mathcal{D} \mathcal{I}\left( \mathcal{D} \right) \|_\infty
\end{equation}
%illustrated in figure \eqref{fig:cond_num}. 
For small $r$, $\frac{\| E \|_\infty}{M}$ is comparable with the 
machine precision $\epsilon = 10^{-16}$; it then increases as $C_2^r$ for some constant $C_2$ and eventually 
saturates around $1$.
In appendix \eqref{appB} a qualitative explanation of the power-law increase of $\frac{\| E \|_\infty}{M}$ is 
provided. 
The gradual corruption of data revealed by the increase of $\| E \|_\infty$ reflects, as illustrated in figure 
\eqref{fig:tyko}, on the quality of the AFQMC estimates of ITCFs, which combine the matrix elements 
of $\mathcal{D}$ and $\mathcal{I}\left(\mathcal{D}\right)$ as prescribed by \eqref{the_boss}.
We propose to mitigate the numeric instability of the ITCF estimator by performing a Tikhonov 
regularization \cite{Tikhonov1977} of the numeric inverse $\mathcal{I}\left( \mathcal{D}\right)$. Practically, 
the SVD of $\mathcal{D}$ is computed:
\begin{equation}
\mathcal{D} = U \mbox{diag}(\sigma_1 \dots \sigma_M) V^\dag
\end{equation}
and $\mathcal{I}\left( \mathcal{D} \right)$ is obtained as:
\begin{equation}
\label{eq:tyko}
\mathcal{I}\left( \mathcal{D} \right) = V \mbox{diag}(\tilde{\sigma}_1 \dots \tilde{\sigma}_M) U^\dag
\end{equation}
where $\tilde{\sigma}_i = \frac{\sigma_i}{\lambda^2 + \sigma_i^2}$ is defined by a regularization parameter 
$\lambda$. Large singular values $\sigma_i \gg \lambda$ are mapped to $\tilde{\sigma}_i \simeq \frac{1}{\sigma_i}$, 
while small singular values $\sigma_i \lesssim \lambda$ are kept below the threshold $\frac{1}{2\lambda}$.
Particular care must be taken in choosing the regularization parameter $\lambda$, since for small $\lambda$ the 
Tikhonov regularization is clearly ineffective, while for large $\lambda$ it provokes a severe alteration 
in $\mathcal{I}\left( \mathcal{D} \right)$. On the other hand, an intermediate value of $\lambda$ prevents 
small errors in $\mathcal{D}$, associated to small singular values $\sigma_i$, to be dramatically amplified 
by the numeric inversion.

The effect of the Tikhonov regularization has been probed considering the model systems of $2$ particles 
introduced in \cite{noi}, for which exact numeric solution of the Hamiltonian eigenvalue problem is feasible, 
and thus the ITCFs is exactly known. 
In figure \eqref{fig:tyko} we show the effect of the Tikhonov regularization \eqref{eq:tyko} on the ITCFs. 
The results show the existence of a broad interval of $\lambda$, comprising the machine precision $\epsilon = 
10^{-16}$, for which the Tikhonov regularization mitigates the numeric instability affecting the AFQMC estimator 
of ITCFs without introducing any appreciable bias besides that introduced by the real local energy and phaseless 
approximations.
The figure displays, in the upper and lower panel respectively, the statistical errors of the 
AFQMC estimations and the discrepancies with respect to the exact results for three different
values of $\lambda$. 
%The calculation with $\lambda=10^{-18}$, under the machine precision,
%coincides with the non-regularized estimation. 
It is evident that, as the imaginary time becomes large, the effect of the 
regularization is very important.  

%%%%%%%%%%%%%%%%%%%%%%%%%%%%%%%%%%%%%%
% RELATIVE TO SIMULATIONS WITH AN 
% EQUAL NUMBER OF STATISTICAL SAMPLES %
%%%%%%%%%%%%%%%%%%%%%%%%%%%%%%%%%%%%%%%

%\begin{figure}
%\includegraphics{regularization3.eps}
%\caption{effetto della regolarizzazione sull'errore che affligge la $F(q,\tau)$ relativa a
%$q = \frac{2\pi}{L} (0,1)$ per un sistema di $N=2$ elettroni ad $r_s=1$; i valori di $\lambda$
%utilizzati sono $\lambda = 10^{-2}, 10^{-4}, 10^{-6}, 10^{-8}, 10^{-10}, 10^{-16} 10^{-17} 10^{-18} e 0$.
%L'errore cresce al decrescere di $\lambda$.}
%\end{figure}

\subsection{Computational Cost}

The AFQMC estimator of ITCFs should join numeric stability and low computational cost. The aim of the present section
is to show that the computational cost of \eqref{the_boss} is $\mathcal{O}(M^3)$, $M$ being the number of orbitals 
constituting the single-particle basis. The contribution $F_w$ to \eqref{the_boss} brought by a single walker of index 
$w$ reads:
\begin{equation}
\label{the_boss2}
\begin{split}
F_w = \sum_{ijklrs} \mathcal{A}_{rs} \, \mathcal{B}_{kl} \, \langle \hat{a}^\dag_r \hat{a}_s \hat{a}^\dag_i \hat{a}_j  \rangle_w
\mathcal{D}_{ik} \mathcal{D}^{-1}_{lj} \\
\end{split}
\end{equation}
where the abbreviation:
\begin{equation}
\langle \cdot \rangle_w = \frac{\langle \Psi^{(w)}_{BP,m}| \cdot |\Psi^{(w)}_{n} \rangle}
{\langle \Psi^{(w)}_{BP,m}|\Psi^{(w)}_n \rangle}
\end{equation}
has been inserted. The generalized Wick's theorem \cite{balian,noi} implies that:
%Despite its cumbersome appearance, \eqref{the_boss2} can be 
%recalling generalized Wick's theorem, according to which:
\begin{equation}
\label{eq:wick}
  \langle \hat{a}^\dag_r \hat{a}_s \hat{a}^\dag_i \hat{a}_j  \rangle_w = 
  \langle \hat{a}^\dag_r \hat{a}_s \rangle_w            \langle \hat{a}^\dag_i \hat{a}_j  \rangle_w
+ \langle \hat{a}^\dag_r \hat{a}_j \rangle_w            \langle \hat{a}_s \hat{a}^\dag_i  \rangle_w
\end{equation}
\eqref{eq:wick} is most conveniently expressed, introducing the definition 
$\mathcal{G}_{ij} = \langle \hat{a}^\dag_i \hat{a}_j  \rangle_w$
and recalling canonical anticommutation relations, as:
\begin{equation}
\label{wick}
\langle \hat{a}^\dag_r \hat{a}_s \hat{a}^\dag_i \hat{a}_j  \rangle_w = 
\mathcal{G}_{rs} \mathcal{G}_{ij} +
\mathcal{G}_{rj} \left( \delta_{is} - \mathcal{G}_{is} \right)
\end{equation}
Combining \eqref{the_boss2} and \eqref{wick} yields:
\begin{equation}
\label{the_boss3}
\begin{split}
F_w &= \sum_{ijklrs} \mathcal{A}_{rs} \, \mathcal{B}_{kl} \, \mathcal{G}_{rs} \mathcal{G}_{ij} \mathcal{D}_{ik} \mathcal{D}^{-1}_{lj} \\
    &+ \sum_{ijklr } \mathcal{A}_{ri} \, \mathcal{B}_{kl} \, \mathcal{G}_{rj}                  \mathcal{D}_{ik} \mathcal{D}^{-1}_{lj} \\
    &- \sum_{ijklrs} \mathcal{A}_{rs} \, \mathcal{B}_{kl} \, \mathcal{G}_{rj} \mathcal{G}_{is} \mathcal{D}_{ik} \mathcal{D}^{-1}_{lj} \\
\end{split}
\end{equation}
Despite its cumbersome appearance, \eqref{the_boss3} can be efficiently evaluated computing the intermediate 
tensors $\mathcal{D}\mathcal{B}$, $\mathcal{A}\mathcal{G}^T$ and $\mathcal{D}^{-1}\mathcal{G}^T$ at the cost 
of $\mathcal{O}(M^3)$ operations, and subsequently computing $F_w$ as:
\begin{equation}
\label{the_boss4}
\begin{split}
F_w &= \left( \sum_r    \left(\mathcal{A}\mathcal{G}^T\right)_{rr} \right) \left( \sum_{il} 
       \left(\mathcal{D}\mathcal{B}  \right)_{il} \left(\mathcal{D}^{-1}\mathcal{G}^T\right)_{li} \right) \\
&+ \sum_{ilr} \left(\mathcal{D}\mathcal{B}\right)_{il} \left(\mathcal{D}^{-1}\mathcal{G}^T\right)_{lr} \mathcal{A}_{ri} \\
&- \sum_{ilr} \left(\mathcal{D}\mathcal{B}\right)_{il} \left(\mathcal{D}^{-1}\mathcal{G}^T\right)_{li} \left(\mathcal{A}\mathcal{G}^T\right)_{ri}
\end{split}
\end{equation}
at the cost of $\mathcal{O}(M^3)$ more operations. The calculation of $F_w$ further simplifies for operators 
$\hat{A}$ whose matrix elements read $\mathcal{A}_{ij} = \mathcal{A}_j \, \delta_{i,a(j)}$ for some function 
$a:\{ 1 \dots M\} \to \{ 1 \dots M\}$. The density fluctuation operator $\rhoq = \sum_{\vett{k}\sigma} \hat{a}
^\dag_{\vett{k}+\vett{q}\sigma} \hat{a}_{\vett{k}\sigma}$ falls within such category.

The complexity $\mathcal{O}(M^3)$ is the best allowed by the phaseless AFQMC methodology: in fact, the calculation 
of ITCFs requires at least $\mathcal{O}(M^3)$ operations to accumulate the matrix $\mathcal{D}$, and the contractions
in \eqref{the_boss3} do not compromise this favorable scaling with the number of single-particle orbitals.

\section{Results}
\label{results}

In the present work we have simulated paramagnetic systems of $N=18,26,42$ electrons at $r_s = 0.1,0.5,1$; 
we show also results for $N=18$ particles at $r_s=2$.
Our calculations qualify the phaseless AFQMC as a practical and useful methodology for the
accurate evaluation of $F(\vett{q},\tau)$, for systems of $N = \mathcal{O}(10^{2})$ correlated fermions
in the continuum.
The complexity scales as $M^3$ ($M$ being the number of basis sets elements), and the absolute 
statistical error of $F(\vett{q},\tau)$ can be kept at the level $10^{-3} - 2.5 \, 10^{-3} $ 
with moderate computational resources even at values of $\tau \simeq 3/ E_F$ for 
$r_s=0.1,0.5,1$ and  $\tau \simeq 2.5 / E_F$ for $r_s=2$, $E_F = 1/r_s^2$ being the Fermi energy. 
 
The number $N$ of electrons constituting the system is inferior to that typically involved in QMC ground state calculations of bulk fermionic systems, but comparable to that used in the context of excited-states calculations through 
imaginary time correlation functions evaluated via configurational QMC methods reported in literature\cite{fc}.

The imaginary time steps used in our calculations were $\dt = 0.003,0.004,0.006,0.008$ $E_{Ha}^{-1}$ at $r_s=0.1,0.5,1,2$ 
respectively.
For each simulation, the number of plane-waves constituting the single-particle Hilbert space has been raised up to $M= 300$ 
according to the number of particles and to the strength of the interaction.
For all calculations, it was verified that decreasing the time step and increasing the number of plane-waves had a 
negligible effect on the ground state energy.
To obtain correct estimates of ITCFs, it is necessary to perform a sufficiently large number
$m$ of backpropagation steps. However, it is well-known \cite{shiwei_bp2} that raising $m$
can result in an increase in variance, which severely limits the possibility of extracting
physical information from the long imaginary-time tails of the ITCFs.
We have used a number of backpropagation steps in the range $m = 200 - 600$. 
When $m=600$ has proved insufficient, to avoid the increases in variance mentioned above, 
AFQMC estimates have been extrapolated to the $m\to\infty$ limit (data obtained by 
extrapolation will be henceforth marked with an asterisk).

\subsection{Imaginary time correlation functions and excitation energies}
\begin{figure*}[t]
\centering
{\huge{\bf{$\mathrm{r_s}=0.1$}}}
$ $

$ $

\includegraphics[width=\textwidth]{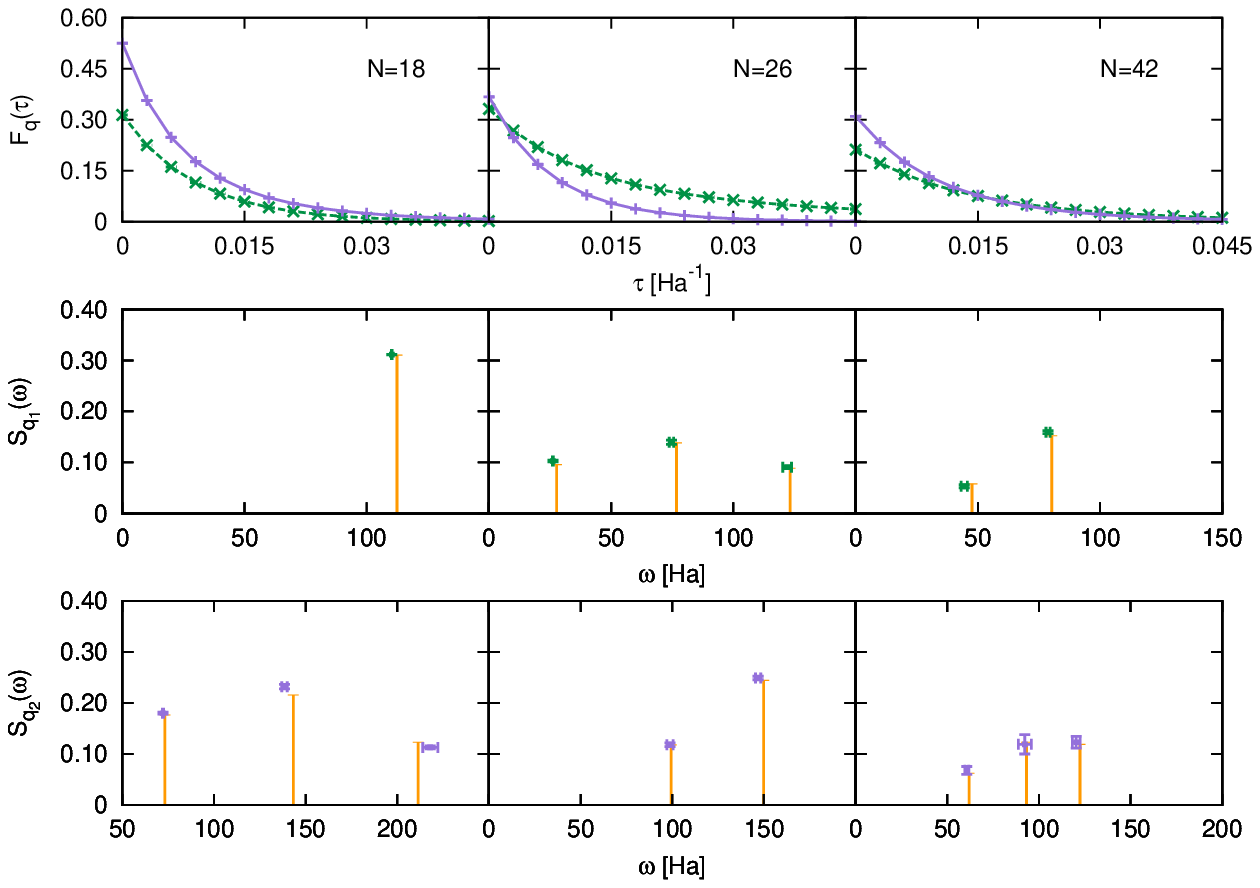}
\caption{(color online) Upper panel: imaginary time correlation functions of the density fluctuation operator 
$\rhoq$ for paramagnetic systems of $N=18,26$ and $42$ particles (left to right) at $r_s=0.1$, with transferred 
momenta $\vett{q}_1$ (green dashed lines) and $\vett{q}_2$ (lavender solid
lines). When not visible, errors are below the symbol size. Lines are only a guide for eyes.
Central panel: dynamical structure factor for $N=18,26$ and $42$ particles (left to right) with 
transferred momentum $\vett{q}_1$ (RPA: orange impulses, AFQMC:
 green symbols).
Lower panel: dynamical structure factor for $N=18,26$ and $42$ particles (left to right) with transferred
momentum $\vett{q}_2$ (RPA: orange impulses, AFQMC: lavender symbols).} \label{fig:rs01}
\end{figure*}

\begin{figure*}[b]
\centering
{\huge{\bf{$\mathrm{r_s}=0.5$}}}
$ $

$ $

\includegraphics[width=\textwidth]{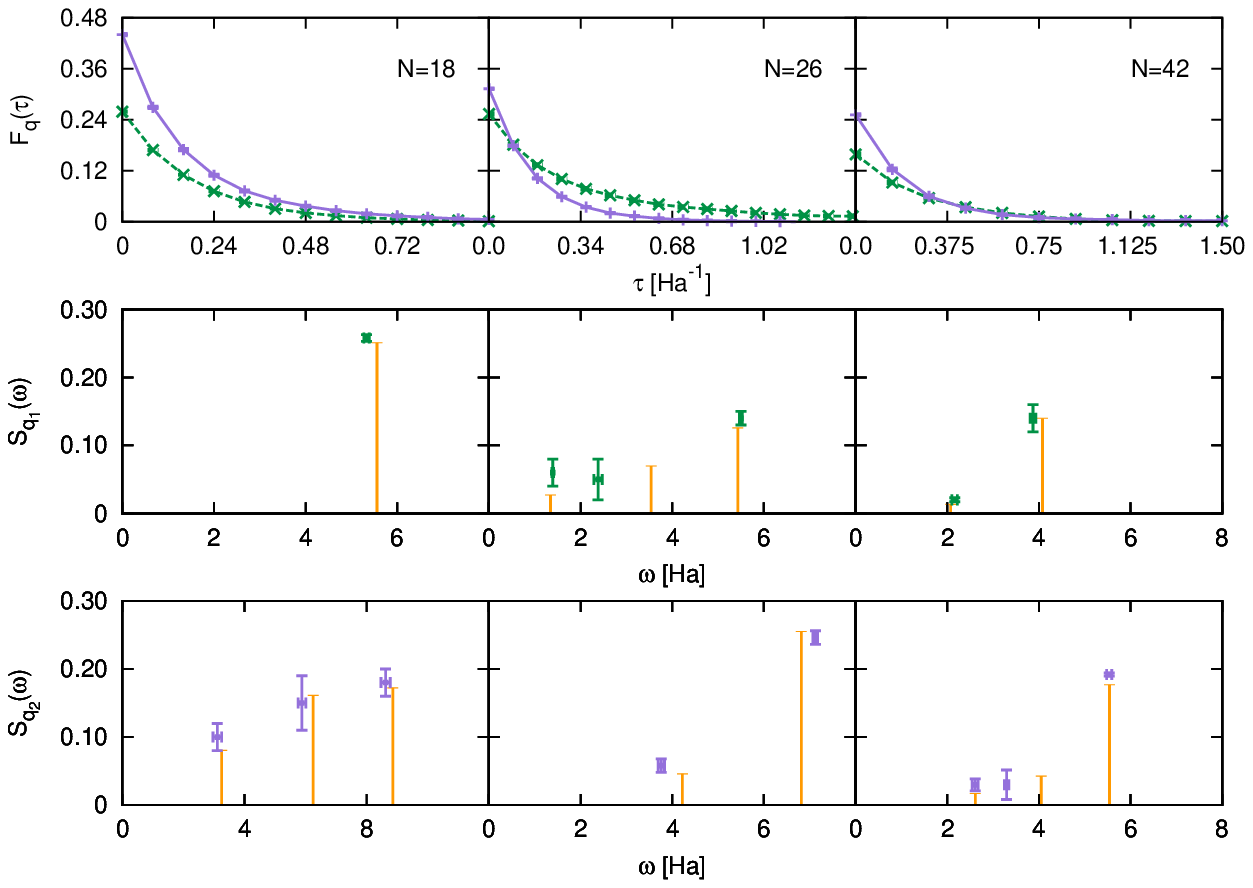}
\caption{(color online) Upper panel: imaginary time correlation functions of the density fluctuation operator $\rhoq$ for 
paramagnetic systems of $N=18,26$ and $42$ particles (left to right) at $r_s=0.5$, with transferred momenta
$\vett{q}_1$ (green dashed lines) and $\vett{q}_2$ (lavender solid lines).
When not visible, errors are below the symbol size. Lines are only a guide for eyes.
Central panel: dynamical structure factor for $N=18,26$ and $42$ particles (left to right) with
transferred momentum $\vett{q}_1$ (RPA: orange impulses, AFQMC:
 green symbols).
Lower panel: dynamical structure factor for $N=18,26$ and $42$ particles (left to right) with transferred
momentum $\vett{q}_2$ (RPA: orange impulses, AFQMC: lavender symbols).
} \label{fig:rs05}
\end{figure*}

\begin{figure*}[ht!]
{\huge{\bf{$\mathrm{r_s}=1$}}}
$ $

$ $

\centering
\includegraphics[width=\textwidth]{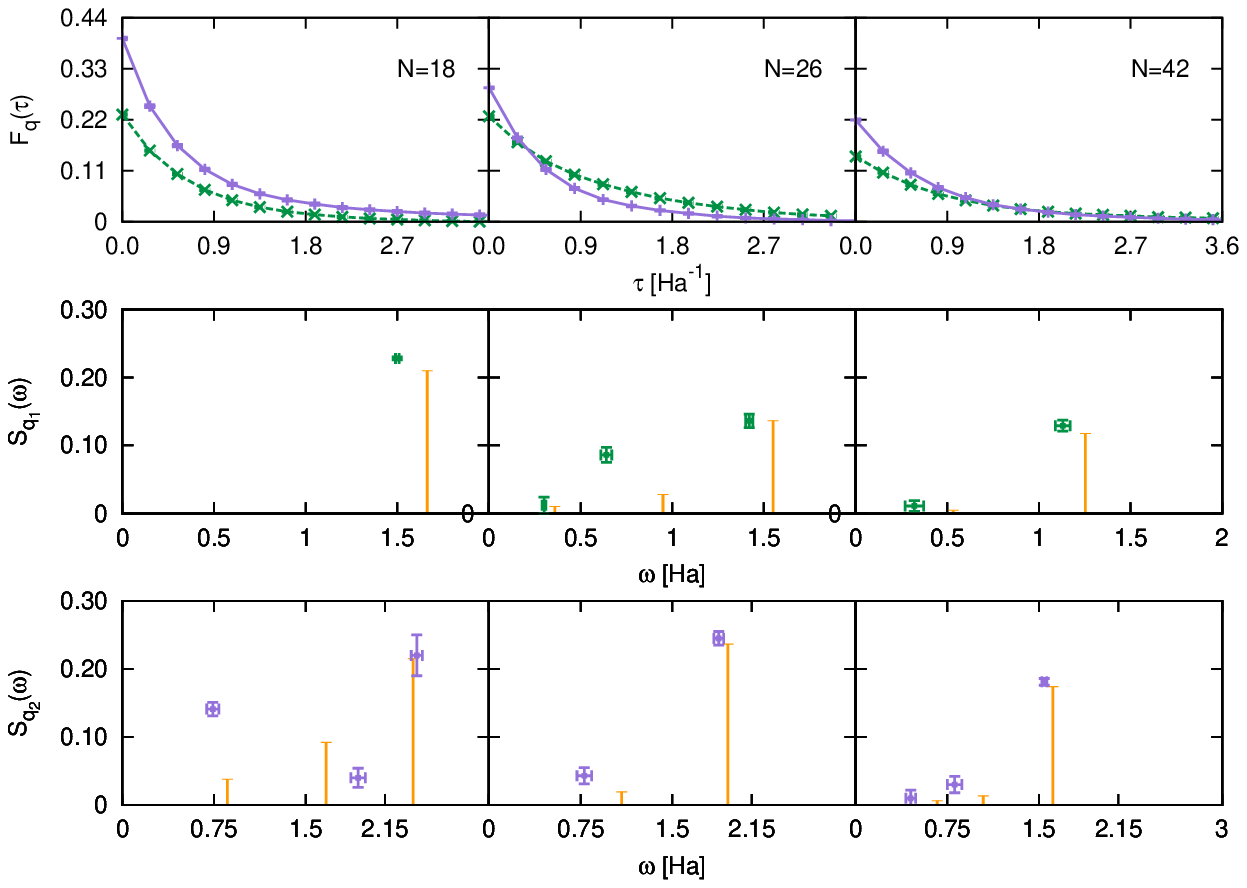}
\caption{(color online) Upper panel: imaginary time correlation functions of the density fluctuation operator $\rhoq$ for
paramagnetic systems of $N=18,26$ and $42$ particles (left to right) at $r_s=  1$, with transferred momenta 
$\vett{q}_1$ (green dashed lines) and $\vett{q}_2$ (lavender solid lines).
When not visible, errors are below the symbol size. Lines are only a guide for eyes.
Central panel: dynamical structure factor for $N=18,26$ and $42$ particles (left to right) with
transferred momentum $\vett{q}_1$ (RPA: orange impulses, AFQMC:
 green symbols).
Lower panel: dynamical structure factor for $N=18,26$ and $42$ particles (left to right) with transferred
momentum $\vett{q}_2$ (RPA: orange impulses, AFQMC: lavender symbols).
} \label{fig:rs1}
\end{figure*}

\begin{figure}[ht!]
{\huge{\bf{$\mathrm{r_s}=2$}}}
$ $

$ $

\centering
\includegraphics[width=0.45\textwidth]{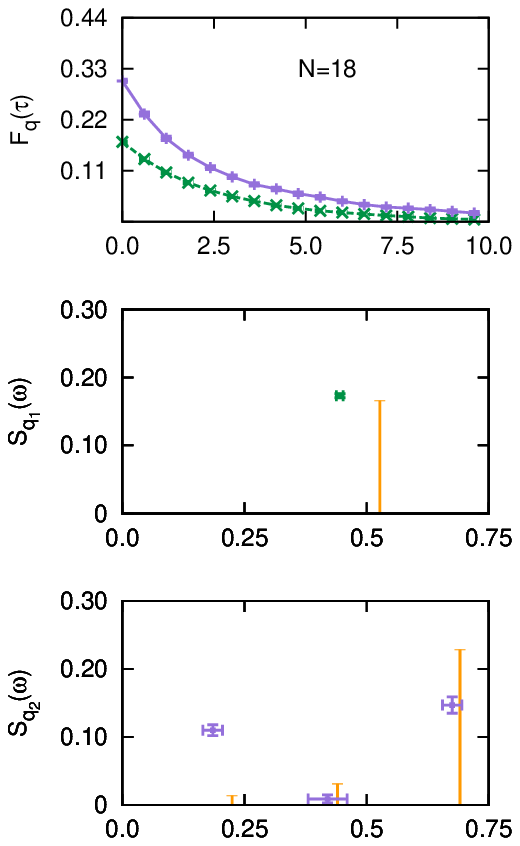}
\caption{(color online) Upper panel: imaginary time correlation functions of the density fluctuation operator $\rhoq$ for
paramagnetic systems of $N=18,26$ and $42$ particles (left to right) at $r_s=  2$, with transferred momenta 
$\vett{q}_1$ (green dashed lines) and $\vett{q}_2$ (lavender solid lines).
When not visible, errors are below the symbol size. Lines are only a guide for eyes.
Central panel: dynamical structure factor for $N=18,26$ and $42$ particles (left to right) with
transferred momentum $\vett{q}_1$ (RPA: orange impulses, AFQMC:
 green symbols).
Lower panel: dynamical structure factor for $N=18,26$ and $42$ particles (left to right) with transferred
momentum $\vett{q}_2$ (RPA: orange impulses, AFQMC: lavender symbols).
} \label{fig:rs2}
\end{figure}

For all the values of $N$ and $r_s$, an AFQMC estimate of the ITCF \eqref{fqt} 
is produced according to the procedure sketched in section \ref{afqmc_dyn_corr}. 
The obtained $F(\vett{q},\tau)$ is shown in the upper panel of Figures~\ref{fig:rs01},~\ref{fig:rs05},~\ref{fig:rs1}~and~\ref{fig:rs2}.
As it is well-known \cite{gift}, it is highly non-trivial to extract physical information
from ITCFs.
In the case of the HEG, the finite size of the systems under study induces to 
expect contributions to $F(\vett{q},\tau)$ coming from excited states of the 
system, which are obtained from the ground state by creation of particle-hole 
pairs. 
The dynamical structure factor $S(\vett{q},\omega)$, the inverse Laplace transform 
of $F(\vett{q},\tau)$, is thus expected to display multiple peaks corresponding to 
the excitation energies. This picture is confirmed by RPA calculations
for finite systems, reported in Appendix \ref{app:RPA}.

The presence of multiple peaks complicates the task of performing the analytic 
continuation providing an estimation of $S(\vett{q},\omega)$.
%Even robust techniques, like the Genetic Inversion via Falsification
%of Theories methodology \cite{gift}, which have provided robust results
%for the dynamical structure factor of Helium systems, are doomed to
%run into difficulties because of the intrinsic features of the
%low-energy excited states of the HEG. 
Therefore, since the number of peaks grows rapidly with the wave-vector modulus 
$|\vett{q}|$, we have limited our attention to the wave-vectors $\vett{q}_1 = (2\pi/L) \, (1,0)$
and $\vett{q}_2 = (2\pi/L) \, (1,1)$.
Notice that $|\vett{q}_1|/k_F = 0.707,0.5,0.447$ and
$|\vett{q}_2|/k_F = 1,0.707,0.632$
for $N=18,26,42$ respectively.
Naturally, $k_F = \sqrt{2}/r_s$ is the Fermi wave-vector.
These low-momentum excitations are very interesting also from a physical point 
of view, in connection with the well-known collective plasmon excitation of the HEG.
Therefore, we do not attempt to predict the full $S(\vett{q},\omega)$,
but limit ourselves to extract the excitations energies and weights by fitting 
the evaluated ITCF to a sum:
\begin{equation}
F(\tau) = \sum_{i=1}^{N_w} s_i e^{-\tau \omega_i}
\end{equation}
%We face the analytical continuation problem 
%
%In this section we present the AFQMC estimates of the dynamical structure factor $S(\vett{q},\omega)$, the Laplace transform of the imaginary time correlation function.
%First, an AFQMC estimate of the ITCF \eqref{fqt} is produced according to the procedure sketched in section 
%\eqref{afqmc_dyn_corr}. 
%To extract information from the estimated ITCF, the latter is fitted to a sum:
%\begin{equation}
%F(\tau) = \sum_{i=1}^{N_w} s_i e^{-\tau \omega_i}
%\end{equation}
of exponential functions with positive frequencies $\omega_i$ and weights $s_i$ 
relying on the well-established Levenberg-Marquardt curve-fitting method
\cite{Levenberg1944}
. The number $N_w \leq 3$ of 
such frequencies and weights is that leading to the best fit.
%\begin{figure*}[ht!]
%\label{fig:fqt2sofq}
%\centering
%\includegraphics[scale=0.55]{gift.eps}
%\caption{Left panel: imaginary time correlation function of the density fluctuation operator $\rhoq$ for a
%paramagnetic system of $N=18$ particles at $r_s=2$, and $\vett{q}= \frac{2\pi}{L} \, (0,1)$.
%Right panel: dynamical structure factor, obtained from the PIFT algorithm.
%}
%\end{figure*}

%\begin{figure}[ht!]
%\label{fig:sofq18}
%\centering
%\includegraphics[scale=0.6]{plas_18p.eps}
%\caption{(color online): color map of normalized $S(\vett{q},\omega)$ for increasing wave vectors at 
%$r_s=0.5,1,2$ (upper, central, lower panel respectively). For increasing the visibility, each 
%$S(\vett{q},\omega)$ has been scaled in order to have maximum value equal to $1$. Energies are measured 
%in units of $\epsilon_F = \frac{1}{r^2_s}$, and wavevectors in units of $k_F = \frac{\sqrt{2}}{r_s}$.}
%\end{figure}

In Figures~\ref{fig:rs01},~\ref{fig:rs05},~\ref{fig:rs1}~and~\ref{fig:rs2} we show results 
relative to the 
simulation of paramagnetic systems at $r_s = 0.1,0.5,1,2$ respectively. Each figure 
contains data relative to the particle numbers $N=18,26,42$ and wave-vectors $\vett{q}_1,\vett{q}_2$.
In the upper panel we show the estimated $F(\vett{q},
\tau)$, while in the middle and lower panels we show, for $\vett{q}_1$ and $\vett{q}_2$
respectively, the obtained frequencies and weights, together with the RPA results.
The AFQMC estimations of the quantities $s_i,\omega_i$ are displayed as points
with both horizontal and vertical statistical errors: the horizontal ones provide the uncertainties
on the frequencies $\omega_i$ of the excitations, while the vertical ones gives the error
bars on the weights $s_i$. The coordinates of the points give, naturally, the mean frequencies and 
weights.
The statistical uncertainties on the quantities $s_i,\omega_i$ are those yield by the fit procedure.
%The uncertainties have been estimated using the jack-knife technique}.
The frequencies predicted by the RPA are represented as impulses
with height equal to the corresponding weights.
We see that, at $r_s = 0.1$, there is a close agreement between AFQMC and RPA predictions of both 
frequencies and spectral weights. Since it is known that, for small $r_s$, RPA predictions are very 
accurate, such agreement provides a robust check for the reliability of AFQMC methodology in 
providing information about the manifold of excited states of the system.
It is well known \cite{noi} that, in the same situation, calculations of $F(\vett{q},\tau)$ based
on the Fixed-Node approximation would give inaccurate results even if the nodal structure of the ground 
state wave function is known with very high accuracy.
As $r_s$ increases, discrepancies appear between the two approaches. The presence of such discrepancies
is naturally expected: none of the methodologies used in the present work is free from approximations.
The approximations underlying RPA and AFQMC, in particular, are quite different in nature and are expected 
to agree only in the limit of high density (very low $r_s$).
Incidentally, we observe that, due to the finite size of the systems investigated,
we cannot provide quantitative predictions about the plasmonic mode, which becomes
a well defined collective excitation in the thermodynamic limit. What we expect is that,
as the system size becomes large enough, of the several peaks that are present for
the finite systems, one will acquire a dominant spectral weight, eventually becoming
a well defined collective excitation. 
In order to further assess the quality of our results, in Tables \ref{tab:eg}, \ref{tab:sq} and 
\ref{tab:chiq} we detail the comparison with the RPA and configurational Monte Carlo results for 
the ground state energy per particle, the static structure factor $S(\vett{q}) = F(\vett{q},0)$ and 
the static density response function:
\begin{equation}
\tilde{\chi}(\vett{q}) = - \frac{\chi(\vett{q})}{2 n} = 
\int_{0}^{+\infty} d\tau \, F(\vett{q},\tau)
\end{equation}
In the AFQMC calculations, $\tilde{\chi}(\vett{q})$ is obtained using the 
parameters yield by the fitting procedure of the density-density 
correlations $F(\vett{q},\tau)$.
As far as ground state energies per particle are concerned, as shown in Table \ref{tab:eg}, at $r_s = 
0.1$ the three methods give compatible results. As $r_s$ increases, 
AFQMC estimates are always closer to FN than RPA, lying between them. 
The configurational Quantum Monte Carlo evaluation of the ground state per particle has been obtained
via Diffusion Quantum Monte Carlo (DMC) calculation, with a nodal structure encompassing backflow 
correlations, optimized by means of the Linear Method \cite{Umrigar2008,Motta2015}.
It is well-known that FN calculations with optimized nodal structures yield 
highly accurate estimates of the ground state energy, as confirmed by comparison with 
Full Configuration Interaction QMC calculations \cite{Alavi2012,Alavi2013}: this result, 
therefore, confirms the great accuracy of the phaseless approximation 
\cite{phaseless1,phaseless2}.

The configurational QMC evaluation of the static structure factor $S(\vett{q})$ has been 
obtained via FN DMC calculations with the nodal structure described above. DMC estimates have been
extrapolated \cite{Foulkes2001}.
Moreover, we have compared AFQMC estimations of the static density response function $\tilde{\chi}(\vett{q})$
with RPA and Fixed-Node estimations.
In principle, the Fixed-Node evaluation of the static density response function $\tilde{\chi}(\vett{q})$ is 
highly non trivial, involving the manifold of excited states. However, it is well-known that this 
difficulty can be circumvented \cite{saverio1} extracting $\tilde{\chi}(\vett{q})$ from the ground state 
energy $E(v_{\vett{q}})$ of a system subject to an external periodic potential of amplitude $v_{\vett{q}}$ 
in the $v_{\vett{q}} \to 0$ limit.
We observe that, increasing $r_s$ above $0.1$, the AFQMC predictions remain, in general, closer to 
the configurational Monte Carlo ones than to the RPA ones: this is a strong indication about the 
quality of AFQMC results, since the Monte Carlo calculations include correlations beyond the RPA level.
This result is remarkable, since the AFQMC evaluation of $\tilde{\chi}(\vett{q})$ is considerably
influenced by the low-energy excitations which, if predicted unaccurately, can significantly bias the result.
%{\color{red} NEI TRE CASI (N=26, rs=0.5 e 1, q = (1,0) e N=18, rs=2, q=(1,1) IN CUI L'ACCORDO CON IL FN E'
%PIU' MEDIOCRE, AVENDO SPINTO LA BACKPROPAGATION FIN DOVE POSSIBILE, STIAMO VERIFICANDO LA POSSIBILITA' 
%DI ESTRARRE $\chi^*(\vett{q})$ PER ESTRAPOLAZIONE IN $n_{bp}$.}
As $r_s$ further increases, however, the agreement decreases. 
%Remarkably, the estimations of 
%$S(\vett{q})$ and $\chi^*(\vett{q})$ have the same trend for all the three methodologies used in the 
%present work.
We have verified that the number of plane-waves and the number of backpropagation steps are 
sufficiently large to extrapolate the results and to filter the excited states contributions from the 
trial wave function.
Hence, the origin of the discrepancies between the estimations yield by the three methodologies used 
in the present work has to be sought in the approximation schemes underlying them.

\begin{table}[ht!]
\begin{tabular}{c c c c c}
\hline\hline \\
$N$ & $r_s$ & $\frac{\epsilon_0}{N}$ (RPA) 
            & $\frac{\epsilon_0}{N}$ (AF) 
            & $\frac{\epsilon_0}{N}$ (FN) \\
\hline \\
18  & 0.1   & 40.14 & 40.14(2)  & 40.13(1) \\
26  & 0.1   & 45.84 & 45.82(1)  & 45.81(1) \\ 
42  & 0.1   & 42.18 & 42.18(1)  & 42.17(1) \\
%\vspace{1pt}
\hline
%\vspace{1pt}
18  & 0.5   & 0.5065 & 0.5007(2)  & 0.5012(2) \\
26  & 0.5   & 0.7520 & 0.7360(2)  & 0.7326(8) \\
42  & 0.5   & 0.6031 & 0.6002(1)  & 0.5922(9) \\
%\vspace{1pt}
\hline
%\vspace{1pt}
18  & 1.0   & -0.2489 & -0.2562(1) & -0.2580(1) \\
26  & 1.0   & -0.1847 & -0.1921(1) & -0.1961(1) \\
42  & 1.0   & -0.2215 & -0.2283(1) & -0.2309(2) \\
%\vspace{1pt}
\hline
%\vspace{1pt}
18  & 2.0   & -0.2661 & -0.2695(1) & -0.2717(1) \\
\hline
\end{tabular}
\caption{RPA (column 3), AFQMC (column 4) and FN-DMC (column 5) estimates of the 
ground state energy for various systems (parameters are listed in columns 1-3); 
energies are measured in $E_{Ha}$.
The RPA ground state energy is calculated on the Gaskell trial wavefunction 
\cite{Gaskell1962}. } \label{tab:eg}
\end{table}

\begin{table}[ht!]
\begin{tabular}{c c c c c c c}
\hline\hline \\
$N$ & $r_s$ & $|\vett{q}|$ & $S(\vett{q})$ (RPA) & 
                             $S(\vett{q})$ (AF)  & 
                             $S(\vett{q})$ (FN) \\
\hline \\
18  & 0.1   & 8.355427     & 0.3105  & 0.314(2) & 0.319(4) \\
18  & 0.1   & 11.81636     & 0.5150  & 0.525(4) & 0.521(4) \\
26  & 0.1   & 6.952136     & 0.3326  & 0.342(2) & 0.343(4) \\
26  & 0.1   & 9.831805     & 0.3623  & 0.367(6) & 0.370(5) \\
42  & 0.1   & 5.469911     & 0.2101  & 0.212(7) & 0.217(4) \\
42  & 0.1   & 7.735622     & 0.3045  & 0.310(6) & 0.306(5) \\
%\vspace{1pt}
\hline 
%\vspace{1pt}
18  & 0.5   & 1.671085     & 0.2511  & 0.258(1) & 0.266(4) \\
18  & 0.5   & 2.363271     & 0.4137  & 0.440(3) & 0.448(5) \\
26  & 0.5   & 1.390427     & 0.2225  & 0.254(3)$^*$ & 0.238(4) \\
26  & 0.5   & 1.966361     & 0.3009  & 0.313(2) & 0.322(4) \\
42  & 0.5   & 1.093982     & 0.1533  & 0.161(2) & 0.146(5) \\
42  & 0.5   & 1.547124     & 0.2366  & 0.247(2) & 0.264(4) \\
%\vspace{1pt}
\hline
%\vspace{1pt}
18  & 1.0   & 0.835543     & 0.2098  & 0.231(2) & 0.218(5) \\ 
18  & 1.0   & 1.181636     & 0.3451  & 0.395(3) & 0.386(4) \\
26  & 1.0   & 0.695214     & 0.1746  & 0.227(2)$^*$ & 0.192(5) \\
26  & 1.0   & 0.983181     & 0.2558  & 0.289(2) & 0.281(4) \\
42  & 1.0   & 0.546991     & 0.1219  & 0.141(1) & 0.126(5) \\
42  & 1.0   & 0.773562     & 0.1938  & 0.219(2) & 0.208(5) \\
%\vspace{1pt}
\hline
%\vspace{1pt}
18  & 2.0   & 0.417771     & 0.1657  & 0.172(2)$^*$ & 0.176(4) \\
18  & 2.0   & 0.590818     & 0.2732  & 0.304(3)$^*$ & 0.305(4) \\
\hline 
\end{tabular}
\caption{RPA (column 4), AFQMC (column 5) and FN-DMC (column 6) 
estimates 
of the static structure factor $S(\vett{q})$ for various systems and 
wave-vectors (parameters are listed in columns 1-3);    
wave-vectors are measured in $a_B^{-1}$. AFQMC estimates
marked
with an asterisk are extrapolated.} \label{tab:sq}
\end{table}

\begin{table}[ht!]
\begin{tabular}{c c c c c c c}
\hline\hline \\
$N$ & $r_s$ & $|\vett{q}|$ & $\tilde{\chi}(\vett{q})$ (RPA) & $\tilde{\chi}(\vett{q})$ (AF) & $\tilde{\chi}(\vett{q})$ (FN) \\
\hline \\ 
18  & 0.1   & 8.355427     & 0.00276                & 0.0028(4)  & 0.00287(1) \\
18  & 0.1   & 11.81636     & 0.00449                & 0.0046(1)  & 0.00469(1) \\
26  & 0.1   & 6.952136     & 0.00598                & 0.0065(2)  & 0.00653(4) \\
26  & 0.1   & 9.831805     & 0.00282                & 0.0028(1)  & 0.00287(4) \\
42  & 0.1   & 5.469911     & 0.00311                & 0.0032(4)  & 0.00325(4) \\
42  & 0.1   & 7.735622     & 0.00330                & 0.0034(2)  & 0.00335(2) \\
\hline
18  & 0.5   & 1.671085     & 0.04516                & 0.048(1)    & 0.0484(4) \\
18  & 0.5   & 2.363271     & 0.06992                & 0.081(2)    & 0.0788(4) \\
26  & 0.5   & 1.390427     & 0.06298                & 0.085(6)$^*$& 0.069(2) \\
26  & 0.5   & 1.966361     & 0.04827                & 0.051(1)    & 0.0504(4) \\
42  & 0.5   & 1.093982     & 0.04074                & 0.043(5)    & 0.042(2) \\
42  & 0.5   & 1.547124     & 0.04903                & 0.051(3)    & 0.049(2) \\
\hline
18  & 1.0   & 0.835543     & 0.12612                & 0.152(3)    & 0.143(2) \\
18  & 1.0   & 1.181636     & 0.18979                & 0.301(6)    & 0.226(1) \\
26  & 1.0   & 0.695214     & 0.14601                & 0.22(2)$^*$ & 0.162(2) \\
26  & 1.0   & 0.983181     & 0.13872                & 0.188(6)    & 0.161(2) \\
42  & 1.0   & 0.546991     & 0.10212                & 0.158(7)    & 0.14(1) \\
42  & 1.0   & 0.773562     & 0.13014                & 0.176(9)    & 0.16(1) \\
\hline
18  & 2.0   & 0.417771     & 0.31451                & 0.34(1)$^*$ & 0.374(4) \\
18  & 2.0   & 0.590818     & 0.46238                & 0.89(1)$^*$ & 0.590(2) \\
\hline
\end{tabular}
\caption{RPA (column 4), AFQMC (column 5) and FN-DMC (column 6) estimates of the compressibility 
$\tilde{\chi}(\vett{q})$ for various systems and wave-vectors (parameters are listed in columns 1-3);
wave-vectors are measured in $a_B^{-1}$, and $\tilde{\chi}(\vett{q})$ in $E_{Ha}^{-1}$. AFQMC estimates marked with an asterisk
are extrapolated.} \label{tab:chiq}
\end{table}

\section{Conclusions}
\label{conclusions}
We have shown the possibility to provide accurate first principles calculations of imaginary time 
correlations for medium-sized fermionic systems in the continuum, using the phaseless Auxiliary Fields 
Quantum Monte Carlo method.

We have simulated a $2D$ homogeneous electron gas of up to $N=42$ electrons using a plane-waves basis set of up
to $M=300$ elements. We have shown that the density-density correlation function in imaginary time can be calculated
via a polynomially complex algorithm with the favorable scaling $\mathcal{O}(M^3)$.
In order to achieve a good accuracy level in the calculations, we propose stabilization procedures to deal with 
matrix inversion, which can be used in combination with well-established stabilization procedures for matrix 
exponentiation and multiplication \cite{Sugar,Alhassid}: in particular, we suggest a Tikhonov regularization that allows 
to maintain a good accuracy level even for imaginary time values of the order of $3/E_F$. We have yielded also comparisons 
with predictions of the static structure factor and the static density response obtained via the RPA approximation and via 
Fixed-Node Quantum Monte Carlo calculations.

At small $r_s$ the AFQMC correctly reproduces the RPA results. At larger $r_s$ on the other hand, it provides 
quantitative estimates of the deviations from the RPA, as the comparison with FN calculations reveals.
We believe this is a relevant result for QMC simulations: it is known, in fact, that the widely employed Fixed-Node 
approximation fails to properly sample the imaginary-time propagator, due to the imposition of the ground-state 
nodal structure to excited states. 
AFQMC, on the other hand, appears to provide a useful tool to explore, from first principles, the manifold of the excited states of a fermionic system.

\section{Acknowledgements}

We acknowledge the CINECA and the Regione Lombardia award, under the LISA initiative, for the availability of
high-performance computing resources and support. 
M. M. also acknowledges funding provided by the Dr. Davide Colosimo Award, celebrating the memory of physicist 
Davide Colosimo.

\appendix

\section{Hubbard-Stratonovich Transformation for the 2DHEG}
\label{appA}

By a straightforward application of the canonical anticommutation relations, the Hamiltonian 
\eqref{heg_ham} can be exactly rewritten as:
\begin{equation}
%\label{heg_ham2}
\ham=\sum_{\vett{k}\sigma} \left( \frac{|\vett{k}|^2}{2} - \mu(\vett{k}) \right) 
                         \,\crt{\vett{k}\sigma}\dst{\vett{k}\sigma} +
\frac{1}{2\Omega} \,\sum_{\vett{q}} \frac{2\pi}{|\vett{q}|} \rhoq \rhomq
\end{equation}
where:
\begin{equation}
\mu(\vett{k}) = \frac{1}{2\Omega} \sum_{\vett{p}\neq\vett{k}} \frac{2 \pi}{|\vett{p}-\vett{k}|}
\end{equation}
and $\rhoq = \sum_{\vett{k}\sigma} \crt{\vett{k}-\vett{q}\sigma}\dst{\vett{k}\sigma}$ is the
density fluctuation operator. Recalling the parity of $\frac{2\pi}{|\vett{q}|}$ and the
anticommutation relation:
\begin{equation}
\left[ \rhoq,\rhomq\right]_+ = \frac{(\rhoq+\rhomq)^2}{2} + \frac{(i\rhoq-i\rhomq)^2}{2}
\end{equation}
one eventually finds:
\begin{equation}
%\label{heg_ham3}
\ham= \ham_0 + 
\frac{1}{2} \,\sum_{\vett{q}} \left(\hat{A}_1(\vett{q})^2 + \hat{A}_2(\vett{q})^2 \right)
\end{equation}
with:
\begin{equation}
\ham_0 = \sum_{\vett{k}\sigma} \left( \frac{|\vett{k}|^2}{2} - \mu(\vett{k}) \right) 
                         \,\crt{\vett{k}\sigma}\dst{\vett{k}\sigma} 
\end{equation}
and:
\begin{equation}
%\label{heg_ham4}
\hat{A}_1(\vett{q}) = \sqrt{\frac{2 \pi}{\Omega |\vett{q}|}} \, \frac{\rhoq+\rhomq}{2}
\quad
\hat{A}_2(\vett{q}) = \sqrt{\frac{2 \pi}{\Omega |\vett{q}|}} \, \frac{i\rhoq-i\rhomq}{2}
\end{equation}
which, since $\rhomq=\rhoq^\dag$, are hermitian operators. Applying the Hubbard-Stratonovich 
transformation to the propagator of the Hamiltonian \eqref{heg_ham3} yields:
\begin{equation}
%\label{hs-transform2}
\hat{G}(\vett{\eta}) = e^{-\frac{\delta\tau}{2} \ham_0}
e^{-i\sqrt{\delta\tau} \sum_{\vett{q}} \eta_{1\vett{q}} \hat{A}_1(\vett{q}) +
                                       \eta_{2\vett{q}} \hat{A}_2(\vett{q}) }
e^{-\frac{\delta\tau}{2} \ham_0}
\end{equation}

\section{Numeric Stability of Matrix Inversion}
\label{appB}

The distance:
\begin{equation}
\label{eq:inv_err}
\| \mathcal{I}\left( \mathcal{D}_r \right) - \mathcal{D}_r^{-1} \|_\infty
\end{equation}
between the actual inverse $\mathcal{D}_r^{-1}$ of $\mathcal{D}_r$ and its numeric estimate $\mathcal{I}\left( \mathcal{D}_r \right)$
\eqref{eq:inv_err} is bounded \cite{Turing1948} by:
\begin{equation}
\label{eq:turing}
\| \mathcal{I}\left( \mathcal{D}_r \right) - \mathcal{D}_r^{-1} \|_\infty 
\leq M
\| \mathcal{I}\left( \mathcal{D}_r \right) \|_\infty \frac{\| E_r \|_\infty}{1 - M \| E_r \|_\infty}
\end{equation}
with:
\begin{equation}
\| E_r \|_\infty = \| \mathbb{I} - \mathcal{D}_r \mathcal{I}\left( \mathcal{D}_r \right) \|_\infty
\end{equation}
Equation \eqref{eq:turing} holds for $\| E_r \|_\infty < \frac{1}{M}$, and is therefore adequate to the description of $\| E_r \|_\infty$
for small $r$. It can be combined with the following estimate \cite{Fox1948}:
\begin{equation}
\label{eq:fox}
\| \mathcal{I}\left( \mathcal{D}_r \right) - \mathcal{D}_r^{-1} \|_\infty \simeq \epsilon 
\| \mathcal{D}_r^{-1} \|^2_\infty \frac{M^3}{3}
\end{equation}
to yield:
\begin{equation}
 M \| E_r \|_\infty \simeq \frac{\epsilon \frac{M^3}{3} \| \mathcal{D}_r^{-1} \|^2_\infty}
                                {  \| \mathcal{I}\left( \mathcal{D}_r \right) \|_\infty + 
                                 \epsilon \frac{M^3}{3} \| \mathcal{D}_r^{-1} \|^2_\infty}
\end{equation}
In the case of AFQMC calculations, where $\mathcal{D}_r$ and $\mathcal{D}_r^{-1}$ come from the product of $r$ matrices:
\begin{equation}
\label{eq:fox2}
\| \mathcal{D}_r^{-1} \|_\infty = C_3^r \quad \| \mathcal{I}(\mathcal{D}_r)\|_\infty = C_4^r
\end{equation}
where $C_3$ and $C_4$ are suitable constants, close to each other. Merging \eqref{eq:fox} and \eqref{eq:fox2}
leads to:
\begin{equation}
\| E_r \|_\infty \simeq \frac{ \epsilon \frac{M^2}{3} }
                             { \left( \frac{C_4}{C_3^2} \right)^r + \epsilon \frac{M^3}{3} }
\end{equation}
which reduces to: 
\begin{equation}
\label{eq:foxeturing}
\| E_r \|_\infty \simeq \epsilon \frac{M^2}{3} \left( \frac{C_3^2}{C_4} \right)^r
\end{equation}
in the limit of small $r$. Since $C_4 < C_3^2$, $C_3$ and $C_4$ being close to each other, the estimate \eqref{eq:foxeturing} 
leads to a power-law increase of $\| E_r \|_\infty$.

\section{RPA for Finite Homogeneous Systems}
\label{app:RPA}

The aim of this appendix is to provide a brief description of the Random Phase
Approximation (RPA) \cite{Sawada1957,fetter,vignale} for finite interacting systems and of the
procedure leading to the excitation energies and weights, with which AFQMC results have been compared.

\begin{figure}[ht!]
\centering
\includegraphics[scale=0.7]{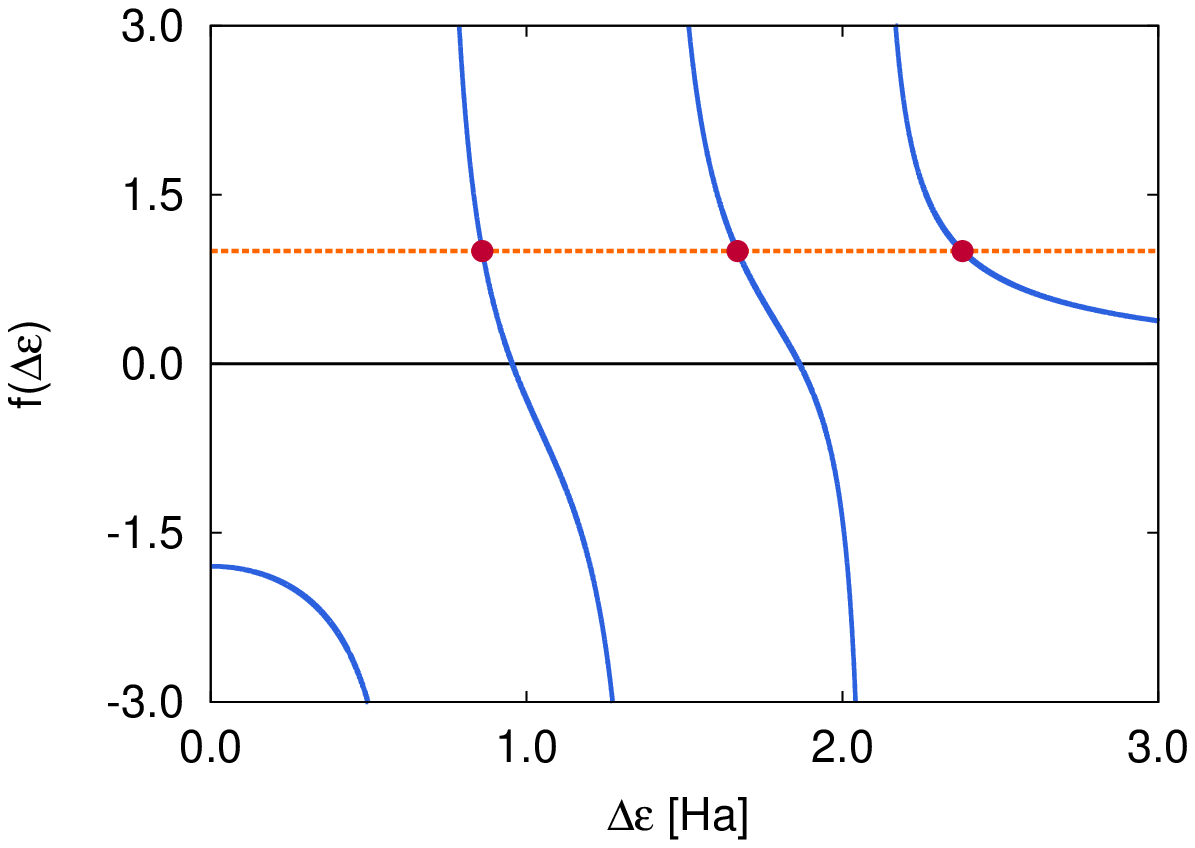}
\caption{(color online)
RPA secular equation for a paramagnetic system of $N=18$ particles at $r_s=1$, and 
$\vett{q}_2$. 
The blue solid line is the right member $f(\Delta\epsilon)$ of \eqref{RPA_sec}, and the 
orange dashed line is the constant function $g(\Delta\epsilon)=1$; intersections between the
two graphs are marked with red dots. the RPA eigenvalues are the abscissae of 
such intersections.
} \label{fig:RPA_sec}
\end{figure}

The RPA can be regarded to \cite{fetter} as a refinement of the well-known Tamm-Dancoff 
approximation \cite{tamm,dancoff,fetter} (TDA), which has long been supporting the study 
of excitations in nuclear systems. 
The TDA relies on the assumptions that the ground state of the system is the Hartree-Fock 
determinant, and that excited states can be represented as superpositions of 
determinants obtained promoting a single particle above the Fermi surface.
Within RPA, on the other hand, a better approximation $\ket{\Phi_0}$ for the actual
ground state of the interacting system is employed to build up an Ansatz for plasmonic 
wavefunctions.
To this purpose, the distinction between spin-orbitals below and above the Fermi level 
is made explicit by writing:
\begin{equation}
\crt{\vett{k}\sigma} = 
\left\{
\begin{split}
\hat{c}^\dag_{\vett{k}\sigma}             \quad \mbox{if $|\vett{k}| > k_F$} \\
\hat{b}^{\phantom{\dag}}_{\vett{k}\sigma} \quad \mbox{if $|\vett{k}| \leq k_F$} \\
\end{split}
\right.
\end{equation}
and the Hamiltonian \eqref{heg_ham} is consequently expressed as:
\begin{equation}
\begin{split}
\ham &= \hat{T} + \hat{V} = \\
     &= \sum_{\vett{k}\sigma} t_{\vett{k}} \, 
\hat{c}^{\dag}_{\vett{k}\sigma}
\hat{c}^{\phantom{\dag}}_{\vett{k}\sigma}
+ 
\sum_{\vett{k}\sigma} t_{\vett{k}} \,
\left( 1 - \hat{b}^\dag_{\vett{k}\sigma}
\hat{b}^{\phantom{\dag}}_{\vett{k}\sigma} \right)
+ \\
&+ \frac{1}{2\Omega} \,\sum_{\vett{q}\neq 0} \phi_{\vett{q}} 
\rhoq
\rhomq
%\rho_{ \vett{q}}
%\rho_{-\vett{q}}
\end{split}
\end{equation}
where $t_{\vett{k}} = \frac{|\vett{k}|^2}{2}$, $\phi_{\vett{q}} = \frac{2\pi}{|\vett{q}|}$,
the first sum goes over all wave-vectors $\vett{k}$ such that $|\vett{k}| > k_F$, the second sum 
goes over all wave-vectors $\vett{k}$ such that $|\vett{k}| \leq k_F$ and the density 
fluctuation operator $\rhoq$ is approximated \cite{Sawada1957} by:
\begin{equation}
\rhoq
%\rho_{ \vett{q}}
\simeq
\sum_{\vett{k}\sigma} 
\hat{c}^{\dag}_{\vett{k+q}\sigma}               \hat{b}^\dag_{\vett{k}\sigma}
+ 
\hat{b}^{\phantom{\dag}}_{\vett{k+q}\sigma}     \hat{c}^{\phantom{\dag}}_{\vett{k}\sigma}
\end{equation}
where the first sum, describing forward scattering processes in which a particle is promoted
above the Fermi level, goes over all wave-vectors $\vett{k}$ such that $|\vett{k}| > k_F$ and 
$|\vett{k}+\vett{q}| \leq k_F$, and the second sum, describing backward scattering processes 
in which a particle is brought back below the Fermi level, goes over all wave-vectors $\vett{k}$ 
such that $|\vett{k}| \leq k_F$ and $|\vett{k}+\vett{q}| > k_F$.
The RPA Ansatz for plasmonic wavefunctions is:
%Within RPA, plasmonic wave-functions are constructed as superposition of forward and backward
%scattering processes
%The RPA identifies plasmonic wave-functions
%Ansatz for plasmonic wave-functions thus reads:
\begin{equation}
\label{RPA_ansatz}
\ket{\Phi_{\vett{q}} } = 
\sum_{\vett{k} \sigma} 
X_{\vett{k}} \hat{c}^{\dag}_{\vett{k+q}\sigma}           \hat{b}^\dag_{\vett{k}\sigma}
\ket{\Phi_0}
+
\sum_{\vett{k} \sigma}
Y_{\vett{k}} 
\hat{b}^{\phantom{\dag}}_{\vett{k+q}\sigma} \hat{c}^{\phantom{\dag}}_{\vett{k}\sigma}
\ket{\Phi_0}
\end{equation}
\eqref{RPA_ansatz} is justified by the observation that the pair destruction operator 
$\hat{b}_{\vett{k+q}\sigma}           \hat{c}_{\vett{k}\sigma}$ 
annihilates the Hartree-Fock determinant but not the actual ground state of the
interacting system.
The eigenvalues $\epsilon$ such that $\ham \ket{\Phi_{\vett{q}} } = \epsilon \ket{\Phi_{\vett{q}} }$
are obtained recalling that the commutators between the Coulomb interaction and the pair
creation and destruction operators can be approximated \cite{Sawada1957} as:
\begin{equation}
[\hat{c}^{\dag}_{\vett{k+q}\sigma}               \hat{b}^\dag_{\vett{k}\sigma},\hat{V}]
\simeq - \frac{\phi_q}{\Omega}
\rhoq
%\rho_{\vett{Q}}
\end{equation}
and:
\begin{equation}
[\hat{b}^{\phantom{\dag}}_{\vett{k+q}\sigma}     \hat{c}^{\phantom{\dag}}_{\vett{k}\sigma},\hat{V}]
\simeq \frac{\phi_q}{\Omega}
\rhoq
%\rho_{\vett{Q}}
\end{equation}
respectively.
Now, since $\ket{\Phi_0}$ and $\ket{\Phi_{\vett{q}} }$ and eigenstates of $\ham$ with eigenvalues
$\epsilon_0$ and $\epsilon = \epsilon_0 + \Delta \epsilon$ respectively, the following identity holds:
\begin{equation}
\begin{split}
0=&\langle \Phi_{\vett{q}} |
(\epsilon - \ham) \, \hat{c}^{\dag}_{\vett{k+q}\sigma} \hat{b}^\dag_{\vett{k}\sigma} | 
\Phi_0 \rangle
= \\
= & \Delta \epsilon \,  
\langle \Phi_{\vett{q}} | \hat{c}^{\dag}_{\vett{k+q}\sigma} \hat{b}^\dag_{\vett{k}\sigma} |
\Phi_0 \rangle
- %\\
\langle \Phi_{\vett{q}} | [\ham, \hat{c}^{\dag}_{\vett{k+q}\sigma} \hat{b}^\dag_{\vett{k}\sigma}]
|
\Phi_0 \rangle 
%= \\
%= &(\epsilon - \epsilon_0 + t_{\vett{k}} - t_{\vett{k+Q}} ) 
%\langle \Phi_{\vett{q}} | \hat{c}^{\dag}_{\vett{k+Q}\sigma} \hat{b}^\dag_{\vett{k}\sigma} |
%\Phi_0 \rangle
%-
%\frac{\phi_q}{\Omega} \langle \Phi_{\vett{q}} | \rho_{\vett{Q}} |
%\Phi_0 \rangle 
\end{split}
\end{equation}
from which:
\begin{equation}
\label{eq:commut1}
\langle \Phi_{\vett{q}} | \hat{c}^{\dag}_{\vett{k+q}\sigma} \hat{b}^\dag_{\vett{k}\sigma} |
\Phi_0 \rangle
=
\frac{ \frac{\phi_q}{\Omega} \langle \Phi_{\vett{q}} | \rhoq |
\Phi_0 \rangle  }{ \Delta \epsilon + t_{\vett{k}} - t_{\vett{k+q}} }
\end{equation}
follows. Similarly:
\begin{equation}
\label{eq:commut2}
\langle \Phi_{\vett{q}} | \hat{b}_{\vett{k+q}\sigma} \hat{c}_{\vett{k}\sigma} |
\Phi_0 \rangle
=
- \frac{ \frac{\phi_q}{\Omega} \langle \Phi_{\vett{q}} | \rhoq |
\Phi_0 \rangle  }{ \Delta \epsilon + t_{\vett{k}} - t_{\vett{k+q}} }
\end{equation}
Equation \eqref{eq:commut1} and \eqref{eq:commut2} can be summed over $\vett{k},\sigma$ 
to yield the secular equation:
\begin{widetext}
\begin{equation}
%\begin{split}
\langle \Phi_{\vett{q}} | \hat{\rho}_{\vett{q}} |
\Phi_0 \rangle
= 
\frac{2 \phi_{\vett{q}}}{\Omega} 
\langle \Phi_{\vett{q}} | \hat{\rho}_{\vett{q}} |
\Phi_0 \rangle %\\
\left(
\sum_{\substack{|\vett{k}| \leq k_F\\|\vett{k+q}| > k_F}}
\frac{1}{ \Delta \epsilon + t_{\vett{k}} - t_{\vett{k+q}} }
-
\sum_{\substack{|\vett{k}| > k_F\\|\vett{k+q}| \leq k_F}}
\frac{1}{ \Delta \epsilon + t_{\vett{k}} - t_{\vett{k+q}} }
\right) %\\
%\end{split}
\end{equation}
\end{widetext}
which, simplifying the matrix element $\langle \Phi_{\vett{q}} | \hat{\rho}_{\vett{q}}
| \Phi_0 \rangle$ in both members, and applying the change of variables $\vett{r} = - 
\vett{k} - \vett{q}$ in the second sum, takes the form: 
\begin{equation}
\label{RPA_sec}
1 = \frac{\phi_{\vett{q}}}{\Omega}
\left[ 2 \, 
\sum_{\substack{|\vett{k}| \leq k_F\\|\vett{k+q}| > k_F}}
\frac{1}{t_{\vett{k}} - t_{\vett{k+q}} + \Delta \epsilon }
+
\frac{1}{t_{\vett{k}} - t_{\vett{k+q}} - \Delta \epsilon } 
\right] 
\end{equation}
where the term between square brakets is immediately identified with the real part
of the $2D$ Lindhard function $\chi_0(\vett{q},\Delta \epsilon)$ \cite{vignale}.
The coefficients $X_{\vett{k}}, Y_{\vett{k}}$ are determined substituting \eqref{RPA_ansatz}
in \eqref{eq:commut1} and \eqref{eq:commut2}, and read:
\begin{equation}
\begin{split}
X_{\vett{k}} &= 
\frac{ \mathcal{N} }{ t_{\vett{k}} - t_{\vett{k}+\vett{q}} + \Delta \epsilon } \\
Y_{\vett{k}} &= - 
\frac{ \mathcal{N} }{ t_{\vett{k}} - t_{\vett{k}+\vett{q}} + \Delta \epsilon } \\
\end{split}
\end{equation}
where $\mathcal{N}$ is a normalization constant. Notice that $X_{\vett{k}}$ is defined 
for $|\vett{k}| \leq k_F$, $|\vett{k+q}| > k_F$ while $Y_{\vett{k}}$ for $|\vett{k}| > k_F$
and $|\vett{k+q}| \leq k_F$.
The right-hand side of \eqref{RPA_sec} is a function $f(\Delta \epsilon) = 
\frac{\phi_{\vett{q}}}{\Omega} \, \chi_0(\vett{q},\Delta \epsilon)$, illustrated in 
Fig.~\ref{fig:RPA_sec}, with the following properties:
\begin{equation}
\begin{split}
\lim_{\Delta \epsilon \to 0} f(\Delta \epsilon) &< 0 \\
\lim_{\Delta \epsilon \to +\infty} f(\Delta \epsilon) &= 0^+ \\
\end{split}
\end{equation}
and diverging in corrispondence to the particle-hole energies 
$t_{\vett{k+q}}-t_{\vett{k}}$.
As a consequence, there exists a root of the secular equation \eqref{RPA_sec} between 
all the poles of $f(\Delta \epsilon)$ and another root above them.
The excited state corresponding to this root has coefficients $X_{\vett{k}}, Y_{\vett{k}}$ 
sharing the same sign, and is therefore a coherent superposition of particle-hole excitations 
describing a collective high-energy oscillation being precursive of the plasmon.
The excited states orresponding to other roots of \eqref{RPA_sec} have coefficients
$X_{\vett{k}}, Y_{\vett{k}}$ with non-constant sign, and therefore take into
account the persistence of non-interacting properties in the spectrum of the 
electron gas, even in presence of Coulomb interaction.

We have seen that the RPA approximation yields an Ansatz for the energies $\epsilon_{\vett{q},n}$
and wavefunctions $\ket{\Phi_{\vett{q},n}}$ of excited states with definite momentum $\vett{q}$, 
which results in the following approximation for the image of the RPA ground state through the 
density fluctuation operator $\rhoq$:
\begin{equation}
\rhoq \ket{\Phi_0} = \sum_n \ket{\Phi_{\vett{q},n}} \braket{\Phi_{\vett{q},n}|\rhoq}{\Phi_0} 
%= \sum_{n,\vett{k}} \ket{\Phi_{\vett{q},n}} \, (X_{\vett{k},n} + Y_{\vett{k},n})
\end{equation}
with:
\begin{equation}
\braket{\Phi_{\vett{q},n}|\rhoq}{\Phi_0}
=
\sum_{\substack{|\vett{k}| \leq k_F\\|\vett{k+q}| > k_F}}
X_{\vett{k},n}
+
\sum_{\substack{|\vett{k}| > k_F\\|\vett{k+q}| \leq k_F}}
Y_{\vett{k},n}
\end{equation}
and for the dynamical structure factor:
\begin{equation}
S(\vett{q},\omega) = \frac{1}{N} \sum_n \delta(\omega-\epsilon_{\vett{q},n}) 
|\braket{\Phi_{\vett{q},n}|\rhoq}{\Phi_0}|^2
%\sum_{\vett{k}} (X_{\vett{k},n} + Y_{\vett{k},n}) |^2
\end{equation}
%Within RPA, it is also possible to compute the ground state momentum distribution $n_{\vett{k}}$,
%given by:
%\begin{equation}
%n_{\vett{k}} = 
%\begin{cases}
%1 - &\frac{1}{2} \sum_{\vett{q},n} | Y^{\vett{q},n}_{\vett{k}}          |^2 \quad |\vett{k}| \leq k_F \\
%    &\frac{1}{2} \sum_{\vett{q},n} | Y^{\vett{q},n}_{\vett{k}-\vett{q}} |^2 \quad |\vett{k}| > k_F \\
%\end{cases}
%\end{equation}


\begin{thebibliography}{99}

\bibitem{wigner} E. P. Wigner, {\em Phys. Rev.} {\bf 46}, 1002 (1934)
\bibitem{bloch} F. Bloch {\em Z. Phys.} {\bf 57}, 549 (1929)
\bibitem{overhauser} A. W. Overhauser, {\em Phys. Rev. Lett.} {\bf 3}, 414 (1959)
\bibitem{ceperley} D. Ceperley, {\em Phys. Rev. B} {\bf 18}, 3126 (1978); D. M. Ceperley and B. J. Alder, {\em Phys. Rev. Lett.} {\bf 45}, 566 (1980)
\bibitem{vignale} G. F. Giuliani and G. Vignale, {\em Quantum Theory of the Electon Liquid}, Cambridge University Press (2005)
\bibitem{shiwei1} S. Zhang and D. Ceperley, {\em Phys. Rev. Lett.} {\bf 100}, 236404 (2008)
\bibitem{Szabo1996}       For a comprehensive review of the existing quantum chemistry methodologies 
                          see for example the book:
                          A. Szabo, and N.S. Ostlund,
                          {\emph{Modern Quantum Chemistry: Introduction to Advanced Electronic
                          Structure Theory}}, Dover Publications (1996)
\bibitem{fetter}          A. L. Fetter and J. D. Walecka, {{Quantum Theory of Many-Particle Systems}}, Dover (2003)
\bibitem{tanatar} B. Tanatar and D. M. Ceperley, {\em Phys. Rev. B} {\bf 39}, 5005 (1989)
\bibitem{kwon} Y. Kwon, D. M. Ceperley and R. M. Martin, {\em Phys. Rev. B} {\bf 48}, 12037 (1993)
\bibitem{saverio1} S. Moroni, D. M. Ceperley and G. Senatore {\em Phys. Rev. Lett.} {\bf 75}, 689 (1995)
\bibitem{sperimentali1} M. Padmanabhan, T. Gokmen, N. C. Bishop, and M. Shayegan, {\em Phys. Rev. Lett.} {\bf 101}, 026402 (2008)
\bibitem{sperimentali2} T. Gokmen, M. Padmanabhan, K. Vakili, E. Tutuc, and M. Shayegan, {\em Phys. Rev. B} {\bf 79}, 195311 (2009)
\bibitem{sperimentali3} Y.-W. Tan, J. Zhu, H. L. Stormer, L. N. Pfeiffer, K. W. Baldwin, and K. W. West, {\em Phys. Rev. Lett.} {\bf 94}, 016405 (2005)
\bibitem{feynman} R. P. Feynman and A. R. Hibbs {\em Quantum Mechanics and Path Integrals}, McGraw-Hill (1965)
\bibitem{loh}    E. Y. Loh et al., {\em Phys. Rev. B} {\bf 41}, 9301 (1990)
\bibitem{fn} J. Anderson, { \em J. Chem. Phys.} {\bf 69}, 1499 (1975)
\bibitem{fn2} P. J. Reynolds, D. M. Ceperley, B. J. Alder and W. A. Lester, { \em J. Chem. Phys.} {\bf 77}, 5593 (1982)
\bibitem{noi} M. Motta, D. E. Galli, S. Moroni and E. Vitali, {\em J. Chem. Phys.} 140, 024107 (2014)
\bibitem{Ceperley1991} D. M. Ceperley, {\em J. Stat. Phys.} {\bf 63}, 1237 (1991)
\bibitem{fci}   G. H. Booth, A. J. W. Thom and A. Alavi {\em J.Chem. Phys} {\bf 131}, 054106 (2009)
\bibitem{assaad} M. Feldbacher and F. F. Assaad {\em Phys. Rev. B} {\bf 63}, 073105 (2001)
\bibitem{jcp_vari} D. M. Ceperley and B. Bernu, J. Chem. Phys. 89, 6316 (1988); P.-M. Zimmerman, J. Toulouse, Z. Zhang, C.-B. Musgrave, C.-J. Umrigar {J. Chem. Phys.} {\bf 131}, 124103 (2009)
\bibitem{fc}    M. Nava, A. Motta, D. E. Galli, E. Vitali and S. Moroni {\em Phys. Rev. B} {\bf 85}, 184401 (2012)
\bibitem{jcp_dyn} G. H. Booth and G. Chan {\em J. Chem. Phys.} {\bf 137}, 191102 (2012)
\bibitem{chan1} G. H. Booth, G. Chan {\em Phys. Rev. B} {\bf 91}, 155107 (2015) 
\bibitem{scalapino} R. Blankenbecler, D. J. Scalapino and R. L. Sugar {\em Phys. Rev. D} {\bf 24}, 2278 (1981)
\bibitem{af1} G.Sugiyama and S. E. Koonin {\em Ann. Phys.} {\bf 168}, 1 (1986)
\bibitem{af3/2} S. Zhang and H. Krakauer, {\em Phys. Rev. Lett.} {\bf 90}, 136401 (2003)
\bibitem{af2} S. Zhang, H. Krakauer, W. A. Al Saidi and M. Suewettana {\em Comp. Phys. Comm.} {\bf 169}, 394 (2005)
\bibitem{senechal} S. Zhang in {\em Theoretical Methods for Strongly Correlated Electron Systems} Springer Verlag (2003)
\bibitem{shiwei_bp} S. Zhang, J. Carlson and J. E. Gubernatis {\em Phys. Rev. B} {\bf 55}, 
7464 (1997)
\bibitem{shiwei_bp2} W. Purwanto and S. Zhang, {\em Phys. Rev. E} {\bf 70}, 056702 (2004)
\bibitem{jcpshiwei1} W. Purwanto, S. Zhang and H. Krakauer {\em J. Chem. Phys.} {\bf 130}, 094107 (2009)
\bibitem{jcpshiwei2} W. Purwanto, H. Krakauer, Y. Virgus and S. Zhang {\em J. Chem. Phys.} {\bf 135}, 164105 (2011)
\bibitem{jcpshiwei3} W. Purwanto, H. Krakauer and S. Zhang {\em Phys. Rev. B} {\bf 80}, 214116 (2009)
\bibitem{Sawada1957} K. Sawada, {\em Phys. Rev.} {\bf 106}, 372 (1957)
\bibitem{rept} S. Baroni and S. Moroni,             {\em Phys. Rev. Lett.} {\bf 82}, 4745 (1999)
\bibitem{Umrigar2008}     J. Toulouse and C. J. Umrigar,
                          \emph{J. Chem. Phys.} 128, 174101 (2008)
\bibitem{Motta2015}        M. Motta, G. Bertaina, D. E. Galli and E. Vitali, {\em Comp. Phys. Comm.} {\bf 190} 62-71 (2015)
\bibitem{ewald} P. P. Ewald {\em Ann. Phys.} {\bf 369}, 253 (1921)
\bibitem{Foulkes2001} W. M. C. Foulkes, L. Mitas, R. J. Needs and G. Rajagopal {\em Rev. Mod. Phys.} {\bf 73}, 33 (2001)
\bibitem{trotter} H. F. Trotter {\em Proc. Amer. Math. Soc.} {\bf 10}, 545 (1959)
\bibitem{suzuki} M. Suzuki {\em Progr. Theor. Phys.} {\bf 56}, 1454 (1976)
\bibitem{hubbard} J. Hubbard, {\em Phys. Rev. Lett.} {\bf 3}, 77 (1959)
\bibitem{stratonovich} R. L. Stratonovich, {\em Sov. Phys. Doklady} {\bf 2}, 416 (1957)
\bibitem{Turing1948} A. M. Turing, {\em Q. J. Mech. Appl. Math.} {\bf 1}, 287 (1948)
\bibitem{Tikhonov1977}    A. N. Tikhonov and V. Y. Arsenin,
                          \emph{Solution of Ill-posed Problems}, Winston \& Sons (1977)
\bibitem{balian}  R. Balian, E. Brezin, {\em Il Nuovo Cimento} {\bf B 64}, 37 (1969)
\bibitem{Gaskell1962} T. Gaskell, {\em Proc. Phys. Soc.} {\bf 77}, 1182 (1961);
T. Gaskell, {\em Proc. Phys. Soc.} {\bf 80}, 1091 (1962)
\bibitem{gift} E. Vitali, M. Rossi, L. Reatto and D. E. Galli {\em Phys. Rev. B} {\bf 82}, 
174510 (2010)
\bibitem{Levenberg1944} K. Levenberg {\em Quart. Appl. Math. } {\bf 2} 164 (1944);
D. Marquardt {\em SIAM J. Appl. Math.} {\bf 11} 431 (1963)
\bibitem{Alavi2012} J. J. Shepherd, G. H. Booth, A. Gr\"uneis, and A. Alavi, {\em Phys. Rev. B} {\bf 85}, 081103(R) (2012)
\bibitem{Alavi2013} J. J. Shepherd, G. H. Booth and A. Alavi {\em J. Chem. Phys.} {\bf 136}, 244101 (2012)
\bibitem{phaseless1} H. Shi, S. Zhang {\em Phys. Rev. B} {\bf 88}, 125132 (2013)
\bibitem{phaseless2} F. Ma, W. Purwanto, S. Zhang, H. Krakauer {\em Phys. Rev. Lett.} {\bf 114}, 226401 (2015)
\bibitem{Sugar} E. Y. Loh Jr., J. E. Gubernatis, R. T. Scalettar, R. L. Sugar and S. R. White, {\em
Interacting Electrons in Reduced Dimensions, NATO ASI Series} {\bf 213} 55-60 (1989)
\bibitem{Alhassid} C. N. Gilbreth and Y. Alhassid, {\em Comp. Phys. Comm.} {\bf 188} 1-6 (2014).
\bibitem{Fox1948} L. Fox, H. D. Huskey and J. H. Wilkinson, {\em Q. J. Mech. Appl. Math.} {\bf 1}, 149 (1948)
%\bibitem{gfmc} M. H. Kalos, {\em Phys. Rev.} { \bf 128}, 1891 (1962)
%\bibitem{cep} M. Boninsegni, and D.M. Ceperley, {\em J. Low Temp. Phys.} {\bf 104}, 339 (1996)
%\bibitem{pigs} A. Sarsa, K.E. Schmidt and W. Magro, {\em J. Chem. Phys.} {\bf 113}, 1366 (2000)
%\bibitem{spigs} D. E. Galli and L. Reatto,          {\em Mol. Phys.} {\bf 101}, 1697 (2003).
%\bibitem{patate} M. Rossi, M. Nava, L. Reatto, and D. E. Galli, {\em J. Chem. Phys.} {\bf 131}, 154108 (2009)
%\bibitem{gift5} M. Jarrell, and J. E. Gubernatis, {\em Phys. Rep.} {\bf 269}, 133 (1996)
%\bibitem{gift8} S. R. White in {\em Computer Simulation Studies in Condensed Matter Physics III}, Springer Verlag (1991)
%\bibitem{gift9} O. F. Syljuasen, {\em Phys. Rev. B} {\bf 78}, 174429 (2008)
%\bibitem{gift10} D. R. Reichman and E. Rabani,{\em J. Chem. Phys.} {\bf 131}, 054502 (2009)
%\bibitem{gift11} A. W. Sandvik Phys. Rev. B 57, 10287 (1998)
%\bibitem{overpress} M. Rossi, E. Vitali, L. Reatto and D. E. Galli, {\em Phys. Rev. B} {\bf 85}, 014525 (2012)
%\bibitem{night} C. J. Umrigar, M. P. Nightingale and K. J. Runge, {\em J. Chem. Phys.} {\bf 99}, 2865 (1993)
\bibitem{tamm} I. Tamm, {\em J. Phys. (USSR)} {\bf 9}, 499 (1945)
\bibitem{dancoff} S. M. Dancoff, {\em Phys. Rev} {\bf 78}, 382 (1950)

\end{thebibliography}
\end{document}